\documentclass[superscriptaddress,showpacs,aps,showkeys
]{revtex4-1}
\usepackage{graphicx}
\usepackage{amsmath}

\begin{document}

\title{\bf Exact relativistic models of thin disks around  static black holes in
a magnetic field}


\author{Antonio C. Guti\'errez-Pi\~neres} 
\email[e-mail:]{gutierrezpac@googlemail.com}
\affiliation{Facultad de Ciencias B\'asicas, Universidad Tecnol\'ogica de
Bol\'ivar, 
CO 131001, Cartagena de Indias, Colombia}
\affiliation{Instituto de Ciencias Nucleares, Universidad Nacional Aut\'onoma de
M\'exico, 
 AP 70543,  M\'exico, DF 04510, Mexico}

\author{Gonzalo Garc\'{\i}a-Reyes} 
\email[e-mail: ]{ggarcia@utp.edu.co}
\affiliation{Departamento de F\'{\i}sica, Universidad Tecnol\'ogica de Pereira,
 A. A. 97, Pereira, Colombia}

\author{Guillermo A. Gonz\'alez} 
\email[e-mail: ]{guillego@uis.edu.co}
\affiliation{Escuela de F\'{\i}sica, Universidad Industrial de Santander, 
A. A. 678, Bucaramanga, Colombia}




\begin{abstract}

The exact superposition of a  central static   black hole
with surrounding  thin disk  in presence of a magnetic field is investigated.
We consider two models of disk,  
one of infinite extension based on a  Kuzmin-Chazy-Curzon metric 
and other finite based on the first Morgan-Morgan disk. We also analyze
 a simple  model of  active galactic nuclei
consisting of  black hole, a Kuzmin-Chazy-Curzon disk and two rods 
representing jets, 
in presence of magnetic field. 
To   explain the  stability of the disks we   consider  the matter of the
disk made of  two pressureless streams of
counterrotating  charged particles (counterrotating model) moving  along
electrogeodesic. Using the Rayleigh criterion we  derivate for circular orbits 
the stability conditions of the particles of the streams.  
The influence of the magnetic field  on the matter properties of the disk and
on its stability are also  analyzed. 
\end{abstract}

\keywords {General relativity;  Einstein-Maxwell equations; exact solutions; thin disks}

\pacs{04.20.-q, 04.20.Jb, 04.40.Nr}

\maketitle

\section{Introduction}

There is a strong observational evidence that active galactic nuclei (AGN),
X-ray transients and gamma-ray bursts (GRBs) are associated with  black holes  
that accrete matter via a surrounding  accretion disk. 
The exact  mechanism by which these phenomena are  produced 
involves the interaction between a rotating black hole, the  accretion disk and
the
electromagnetic field \cite{Kundt,Krolik,Blandford}. However,  
magnetic fields  play a key role in  understanding   of these process. 
The enormous  observed energy in  AGN
is related with the presence of magnetic field in these nuclei 
\cite{BZ,Zakharov,Silantev}. 
The existence of radio jets is also attributed to  the presence of strong
magnetic fields in
centers of AGN and quasars \cite{Liu, Akiyama, Vlemming, Meier, Ustyugova,
Krasnopolsky}. 

A general exact relativistic model that describes such  astrophysical objects  
require an exact solution of coupled  Einstein-Maxwell field equations  that
represent the 
superposition of a Kerr black hole with a stationary disk and electromagnetic
fields. As a first approximation one could consider  
a static  system composite by a  Schwarzschild  black hole and a 
thin  disk immersed in  a magnetic field.  Exact solutions of the Einstein
equations representing the field of a 	
static thin disks without radial pressure
were first studied by Bonnor and Sackfield \cite{BS}, 
and Morgan and Morgan \cite{MM1}, and with radial  pressure by Morgan and Morgan
\cite{MM2}. 
Several classes of exact solutions of the  Einstein field  equations
corresponding to static thin disks with or  without radial pressure have been
obtained by different authors 
\cite{LP,CHGS,LO,LEM,BLK,BLP,GE}.
Rotating  thin disks that can be considered as a source of a Kerr metric were
presented by  Bi\u{c}\'ak and  Ledvinka \cite{BL}, while rotating disks with
heat flow were were studied by Gonz\'alez and Letelier \cite{GL2}.
The static superposition of a disk and a black hole was first considered
by Lemos and Letelier \cite{LL1,LL2,LV-AGN}.
Thin disks in presence of electromagnetic field  have been discussed as
sources
for Kerr-Newman fields \cite{LBZ}, conformastationary metrics \cite{KBL} 
and for  magnetostatic  axisymmetric fields in  \cite{LET1,GG1,GG-CRIS,Guti1}. 
Monopole and dipole layers in curved spacetimes
were analyzed in \cite{Guti2}, and electromagnetic sources distributed on shells 
in a Schwarzschild background in \cite{Guti3}. Also, conformastatic disk-haloes in Einstein-Maxwell gravity  
were considered in \cite{AN-GUI-QUE},
variational thermodynamics of relativistic thin disks in \cite{AN-QUE}, and
models of perfect fluid disks in a  magnetic field for
conformastatic spacetimes in \cite{GON-OMAR}.
 
In this work we consider the exact static superposition of a  Schwarzschild 
black hole
and  a thin disk  in presence of a magnetic field.  
The method used to include  the magnetic field   is the well-known   complex
potential formalism proposed by  Ernst \cite{E1,E2}, using as seed solutions 
simple vacuum  spacetimes  
representing  the field of a thin disk  and a black hole.

The paper is organized as follows. In Sec. II we discuss the Einstein-Maxwell
equations in the case of magnetostatic  fields and   we present a
summary of the procedure to obtain  models of  thin disks  with a
purely azimuthal pressure  and  current.
In order to have a  stable configuration  in absence of radial pressure, 
the matter in the disks also in interpreted as made  
of  two pressureless (dust) streams of  counterrotating charged particles
(counterrotating model) moving  along electrogeodesic.  
Using the Rayleigh criterion we  derivate for circular orbits 
the stability conditions of the particles of both streams. 

In Sec. III  the formalism for superposing the field of a disk and a static  black hole in the vacuum
\cite{LL2}
is extended to the case of  magnetized Weyl fields.
In Sec. IV  we consider two models of disks,  
one of infinite extension based on a  Kuzmin-Chazy-Curzon metric 
and other finite based on the first Morgan-Morgan disk.
Also  a simple  model of  active galactic nuclei is studied
based on the superposition  of  a black hole, a Kuzmin-Chazy-Curzon disk and two rods 
representing jets \cite{LV-AGN},  in presence of magnetic field.
Finally, in Sec. V we summarize and discuss the results obtained.


\section{Einstein-Maxwell equations  and disks } 

The line element  for a static axially symmetric spacetime in   Weyl's canonical
coordinates
$(t, \varphi,\rho, z)$ is given by \cite{KRAMER}
\begin{equation}
ds^2 = - \ e^{2 \psi} dt^2 \ + \ e^{- 2 \psi} [\rho^2 d\varphi^2 + e^{2 \Lambda}
(d\rho^2 + dz^2)] , \label{eq:met}
\end{equation}
where  $\psi$ and $\Lambda$  are functions of the coordinates $\rho$ and $z$
only. 
The vacuum Einstein-Maxwell equations, in  geometrized units such that 
$ G = c  = 1$,  are given by 
\begin{subequations}\begin{eqnarray}
&   &    R_{ab} \  =  \ 8 \pi T_{ab}, \label{eq:einmax1}  \\
&   &   \nabla_b F^{ab} = 0,  \label{eq:einmax2}
\end{eqnarray}\end{subequations} 
where 
\begin{equation}
T_{ab} \  =  \ \frac{1}{4 \pi} \left [ F_{ac}F_b^{ \ c} - \frac 14
g_{ab}F_{cd}F^{cd} \right ] \label{eq:tab} 
\end{equation}
is the electromagnetic energy-momentum  tensor,
\begin{equation}
F_{ab} =  A_{b,a} -  A_{a,b} 
\end{equation}
the electromagnetic field tensor, and $A_a$  the  electromagnetic four
potential. 
For magnetostatic  axially symmetric fields    $A_a=\delta_a ^\varphi A$, 
where  $A$ is the magnetic potential which  also is function of $r$ and $z$
only. 
The other symbols have the usual meaning, i.e.,  $( \ )_{,a}=\partial /\partial
x^a$,  $\nabla_b$   covariant derivate, etc. 

For the metric (\ref{eq:met}), magnetostatic  Einstein-Maxwell
equations give 
\begin{subequations}\begin{eqnarray}
\nabla\cdot[\rho^{-2}f\nabla A]&=&0, \label{eq:e-m1} \\
f\nabla^{2}f&=&\nabla f\cdot\nabla f + 2\rho^{-2}f^{3}\nabla A\cdot\nabla A,
\label{eq:e-m2} \\
\Lambda,_\rho&=&\rho\left(\psi^{2},_{\rho}-\psi^{2},_{z}\right)+\frac{1}{\rho}
\left(A^{2},_{\rho}-A^{2},_z\right)f, \label{eq:e-m3} \\
\Lambda,_z&=&2\rho\psi,_{\rho} \psi,_{z}+\frac{2}{\rho}A,_{\rho}A,_{z}f,
\label{eq:e-m4}
\end{eqnarray}\end{subequations}
where  $\nabla $  is the standard differential operator in cylindrical
coordinates and $f=e^{2\psi}$. In vacuum, the solutions of the  above system of
equations correspond
to the well-known  Weyl solutions ($\hat \psi$, $\hat \Lambda$) and the equation 
(\ref{eq:e-m2}) is the Laplace's equation. 

Solutions of the Einstein-Maxwell equations (\ref{eq:einmax1}) - (\ref
{eq:einmax2})  representing the field of a thin disk at $z=0$ with electric
current
can be constructed assuming the components of the metric tensor and 
the electromagnetic potential  continuous across the disk, and its first
derivatives discontinuous in the direction normal to the  disk.  This can be
written  as 
\begin{subequations}\begin{eqnarray}
b_{ab} \ &=& [g_{ab,z}] =  g_{ab,z}|_{_{z = 0^+}} \ - \ g_{ab,z}|_{_{z = 0^-}}
 \ = \ 2 \ g_{ab,z}|_{_{z = 0^+}}, \label{eq:b}  \\
 a_{b} \ &=&  [A_{b,z}] =  A_{b,z}|_{_{z = 0^+}} \ - \ A_{b,z}|_{_{z = 0^-}}
 \ = \ 2 \ A_{b,z}|_{_{z = 0^+}}.    \label{eq:a}              
\end{eqnarray}\end{subequations}

The application of the formalism of distributions in curved spacetimes to
the Einstein-Maxwell equations \cite{PH,LICH,TAUB,IS1,IS2} give us
\begin{subequations}\begin{eqnarray}
R_{ab} & = &  8 \pi T_{ab} , \label{eq:EMdisk}  \\
T_{ab} & = &  T^{\text{elm}}_{ab} + T^{\text{mat}}_{ab}  
= T^{\text{elm}}_{ab} +  Q_{ab} \ \delta (z) , \label{eq:Tabdisk}  \\
 \nabla_b F^{ab} & = & 4 \pi J^a ,  \\
J^a & = & j^a \delta (z) ,
\end{eqnarray}\end{subequations}
where $\delta(z)$ is the  usual Dirac function with support on  the disk, 
$T^{\mathrm {elm}}_{ab}$ is the electromagnetic tensor (\ref{eq:tab}),
\begin{equation}
Q^a_b = \frac{1}{16 \pi}\{b^{az}\delta^z_b - b^{zz}\delta^a_b +  g^{az}b^z_b -
g^{zz}b^a_b + b^c_c (g^{zz}\delta^a_b - g^{az}\delta^z_b)\}  \label{eq:Qab} 
\end{equation}
is the  energy-momentum tensor on plane $z=0$, and 
\begin{equation}
j^a =  \frac{1}{4 \pi} [F^{ab} ] \delta^z_b, \label{eq:ja} 
\end{equation}
is the electric current density on the disk. $[F^{ab} ]$ means the jump of
Maxwell tensor across the disk.  
The ``true''  surface
energy-momentum tensor  of the  disk $S_{ab}$ and the ``true'' surface
current density $\text{\sl j}_a$ are given by 
\begin{subequations} \begin {eqnarray}
S_{ab} & = & \int T^{\mathrm {mat}}_{ab} \ ds_n =  \sqrt{g_{zz}} Q_{ab}  ,   \label{eq:Sab}  \\ 
\text{\sl j}_a  & = & \int J_{a}  \ ds_n = \sqrt{g_{zz}} j_a ,  \label{eq:jatrue}  
\end{eqnarray}\end{subequations}
where $ds_n = \sqrt{g_{zz}} \ dz$ is the ``physical  measure'' of length in the
direction normal to the disk.  For the metric (\ref{eq:met}), the nonzero
components of  $S_a^b$ are
\begin{subequations}\begin{eqnarray}
&S^t_t &= \ \frac{1}{4 \pi} e^{\psi - \Lambda} \left\{ \Lambda,_z - \ 2 \psi,_z  \right\} ,
 \label{eq:emt1}     		\\
&	&	\nonumber	\\
&S^\varphi_\varphi &= \ \frac{1}{4 \pi} e^{\psi - \Lambda} \Lambda,_z , \label{eq:emt2}
\end{eqnarray}\label{eq:emt}\end{subequations}
and in the magnetostatic  case the only  nonzero component of the current
density
is  
\begin{equation}
\text{\sl j}_{\varphi} = \ - \frac{1}{2 \pi} e^{\psi - \Lambda} 
A _{,z},  \label{eq:jacimut} 
\end{equation}
where all the quantities are evaluated at $z = 0^+$.

In terms of the orthonormal  tetrad ${{\rm e}_{ (a)}}^b = \{
V^b , W^b , X^b,Y^b \}$, where
\begin{subequations}\begin{eqnarray}
V^a &=& e^{- \psi} \ ( 1, 0, 0, 0 ) , \quad  	
W^a = \frac{e^\psi} {\rho} \ \ ( 0, 1, 0, 0 ) , \label{eq:tetrad1}	\\
X^a &=& e^{\psi - \Lambda} ( 0, 0, 1, 0 ) , \quad
Y^a = e^{\psi - \Lambda} ( 0, 0, 0, 1 ) , \label{eq:tetrad2}
\end{eqnarray}\end{subequations}
the surface energy density $\epsilon$ and  the  
azimuthal pressure $p_\varphi$ on the disk are given by
\begin{equation}
\epsilon \ = \ - S^t_t , \quad p_\varphi \ = \ S^\varphi_\varphi , 
\label{eq:enpr}
\end{equation}
and  the azimuthal current density \text{\sl j} by
\begin{equation}
\text{\sl j}  = W^\varphi \text{\sl j}_\varphi \label{eq:j}.
\end{equation}


Thus  we have a disk  only with pressure and electric current in azimuthal
direction. 
Because there is no radial
pressure or tension to support the gravitational attraction, the matter
distribution  is unstable.  In addition,  
since the spacetime  is static we have no rotation.
In order to have a  stable configuration 
in absence of radial pressure, we need 
assume the counterrotating hypothesis, that is  the matter in the disk is
considered made    of  two
pressureless  streams of  counterrotating charged particles, i.e., that
circulate in opposite directions.
Even though this interpretation
can be seen as merely theoretical, there are observational
evidence of  counterrotating matter components  in certain types of galaxies
\cite{RGK,RFF,BER,STRUCK,CBG}.  We assume 
\begin{subequations}\begin{eqnarray}
S^{ab} &=& S_+^{ab} \ + \ S_-^{ab} \ , \label{eq:Sabsum}   \\
\text{\sl j}^a    &=& \text{\sl j}_+^a + \text{\sl j}_-^a,  \label {eq:jsum} 
\end{eqnarray}\end{subequations}
where
\begin{subequations}\begin{eqnarray}
S^{ab} _\pm &=& \epsilon_\pm \ u_\pm ^a u_\pm ^b, \label{eq:sabcon} \\
\text{\sl j}^a_\pm  &=& \sigma _\pm u_\pm ^a.  \label{eq:jacon}
\end{eqnarray}\end{subequations}
$\epsilon _\pm$  are the matter densities of each stream,  $\sigma_\pm$  the
electric charge  densities, and $u_\pm^a$  the   normalized four-velocities
($u_\pm^a u_{a \pm }
=-1$), which for circular orbit are  $(u_\pm^a) = ( u_\pm^0 , u_\pm^1, 0 , 0 )=
u_\pm^0(1, \omega _\pm, 0 , 0 )$,
where  $\omega_\pm = u_\pm^1/u_\pm^0$ are the angular velocities of each
stream. Now,  using the   continuity equation (the Bianchi identity)   
$T^{ab}_{\ \ \ ;b}=0$ and   Eq. (\ref{eq:Tabdisk})  $T^{ab} =  T^{ab}_{elm} +  T^{ab}_{mat} $, 
we obtain     $T^{ab}_{mat;b} = -T^{ab}_{elm;b}$.
But from  Maxwell equations  $T^{ab}_{elm;b} = - F^a_{\ \ c} J^c$ , then    $T^{ab}_{mat;b} =  F^a_{\ \ c} J^c$. 
In  the disk  this gives  $S^{ab}_{\ \ ;b} =  F^a_{\ \ c}  \text{\sl j}^c$ and   
using (\ref{eq:Sabsum}) and  (\ref{eq:jsum}) we have  $S^{ab}_{\pm \ \ ;b} = F^a_{\ \ c} \text{\sl j}^c_\pm$.
Thus using  (\ref{eq:sabcon}) and  (\ref{eq:jacon}) we obtain
\begin{equation}
\epsilon _\pm u^b_\pm u^a_{\pm ; b} = \sigma_\pm F^a_{\ \ b} u^b_\pm,   
\end{equation}
i. e., each stream follow a electrogeodesic motion.  For  circular orbits  reads 
\begin{equation}
\frac 12 \epsilon _\pm g_{ab,\rho}u^a_\pm u^b_\pm = - \sigma _\pm F_{\rho a}
u^a_\pm,
\label{eq:elecgeo}
\end{equation}
and   in the  case magnetostatic 
\begin{equation}
 \frac 12  \epsilon _\pm u^0_\pm (g_{\varphi \varphi,\rho } \omega_\pm ^2 + g_{t
t,\rho } )= - \sigma_\pm A_{,\rho} \omega_\pm,  \label{eq:motion}
\end{equation}
where $u^0_\pm$ obtains normalizing  $u^a_\pm$,  that is  
\begin{equation}
 (u^0_\pm)^2 = - \frac {1}{g_{\varphi \varphi} \omega_\pm^2 + g_{tt}}. 
\label{eq:u0}
\end{equation}

Using (\ref{eq:Sabsum}) - (\ref{eq:jacon}) and (\ref{eq:motion}) one finds \cite{{KLE3}}-\cite{GGR-Inter}
\begin{subequations}\begin{eqnarray}
\epsilon_\pm  (u^0_\pm)^2 & = & \frac{-  \omega _\mp S^{tt} }{\omega_\pm -  \omega _\mp}, \\
\sigma_\pm u^0_\pm  & = &  \frac{\text{\sl j}^{\varphi}}{\omega_\pm -  \omega _\mp},  \\
\omega_ \pm ^2  &=& -\frac{g_{tt,\rho}}{ g_{\varphi \varphi,\rho} +2 A_{,\rho} \text{\sl j}^{\varphi}
/ S^{\varphi \varphi} }.  \label{eq:omegacontra}  
\end{eqnarray}\end{subequations}

With respect to the  orthonormal tetrad (\ref{eq:tetrad1}) - (\ref{eq:tetrad2})
the 3-velocity has components
\begin{equation}
 v_\pm ^{ (i)}=  \frac { {{\rm e}^{ (i)}}_a u_\pm ^a } { {{\rm e}^{(0)}}_b
u_\pm^b } ,              
\end{equation}
and for equatorial circular  orbits  the 
only nonvanishing velocity components is given by 
\begin{equation}
 (v_\pm^{ (\varphi)})^2 = v^2= - \frac{ g_{\varphi \varphi} }{ g_{tt} } \omega
_\pm^2 = \rho ^2 {\rm e}^{-4\psi} 
\omega _\pm^2   ,  \label{eq:vc2}
\end{equation}
which  represents the circular speed  of the particles as seen by an observer
at infinity. In vacuum the speed  $v$ of counterrotation (rotation curves or rotation
profile) of the particles in the disk is given by 
\begin{equation}
v^2=\frac{p_\varphi}{\epsilon}.
\end{equation}


To analyze the stability of  the particles of the two streams in the case of
circular orbits in the equatorial plane we
use  an extension of Rayleigh  criteria of stability of a fluid at rest
in a gravitational field  \cite{RAYL,FLU,LETSTAB}.
The  method works as follows.   Any small element 
of the matter distribution  analyzed  (in our case  a test particle of the 
streams) is  displaced  slightly   from its path.  
As a result of this displacement, forces appear which
act on the displaced matter  element.  If the matter distribution is stable,
these forces must tend to return the element
to its original position.

The  relativistic Lagrangian for a test  particle of the streams  in presence of a
gravitational and magnetic field is given by 
\begin{equation}
{\cal L_\pm} = \frac 12 g_{ab} u ^a_\pm  u^b_\pm + \tilde  e_\pm A u^ 
\varphi _\pm,
\end{equation}
where $\tilde e _\pm$ is the specific electric charge. 
For magnetostatic  axially symmetric fields  there are  two constants of motion 
\begin{eqnarray}
E_\pm &= &  - g_{tt} u^0_\pm, \\
L _\pm & =&   g_{\varphi \varphi } \omega_\pm u^0_\pm  +  \tilde e _\pm A ,
 \label{eq:L}
\end{eqnarray}
where $E_\pm$ is the relativistic specific energy and $L_\pm$ the specific 
angular momentum. 
The motion equation (\ref{eq:motion})  can be cast as a balance equation
\begin{equation}
\frac {  g^{\rho \rho}  g_{tt,\rho} E_\pm^2 } {2 g_{ tt}^2 }
+  e g^{\rho \rho} A_{,\rho} \omega _\pm u_\pm^0 
 = - \frac { g^{\rho \rho}  g_{\varphi \varphi,\rho} (L_\pm -  \tilde e_\pm
 A)^2 } 
{ 2 g_{\varphi \varphi}^2 }  \label{eq:equilibrio}
\end{equation}
where the term first on the left-hand side represents the gravitational force
$F_g$, the term second  the the Lorentz force $F_L$,  and the term on the
right-hand side  the centrifugal force $F_c(\rho)=F(\rho,L_\pm(\rho))$  acting
on
the test particle. So  we have a balance between the total force $F(\rho) =F_g +
F_L$
and the centrifugal force.   We now
consider the particle to be initially in a circular orbit with 
radius $\rho=\rho_0$ and we  slightly displace it to a higher orbit
$\rho>\rho_0$. The angular momentum of particle  remains equal to its initial
value  $L_{\pm 0}=L_\pm(\rho_0)$ which implies that the centrifugal force   in
its new
position is $ F_c(\rho,L_{\pm 0})$. In order that the particle returns to it
initial
position must be met that  $F(\rho)>F_c(\rho,L_{\pm 0})$, but 
according to the balance equation (\ref{eq:equilibrio})
$F(\rho)=F_c(\rho,L_{\pm})$ so
that
$F_c(\rho,L_{\pm}) > F_c(\rho,L_{\pm 0})$,  and  hence $(L_{\pm} -  \tilde e _\pm
A)^2 > (L_{\pm0} -  \tilde e _\pm
A)^2 $. Using the expression for $L_\pm$ (\ref{eq:L})  and defining 
the function $h_{\pm}=g_{\varphi \varphi } \omega _{\pm} u^0_{\pm}$,
follows that  $h_{\pm}(\rho)^2>h_{\pm}(\rho_0)^2$. The quantity $h_{\pm}$ can be
written as
\begin{equation}
 h_ \pm =  \frac{ \rho e^{-\psi} v} {\sqrt{1 - v^2}},
\end{equation}
and   has the same
form that the specific angular momentum in the vacuum, being  the 
true specific angular momentum  corresponding to expression  (\ref{eq:L}). By
doing a Taylor
expansion of
$h^2_{\pm}(\rho)$ around $\rho=\rho_0$ one finds that  the condition of
stability for  equatorial circular orbits  is   
\begin{equation}
h_{\pm} h_{_{\pm},\rho}  >0,
\end{equation}
or, in other words, $ h_{_{\pm},\rho}^2  >0$. 
Thus when the counterrotating hypothesis is assumed, the stability of the disks 
is equivalent to the stability of the particles of the two streams.


\section{Superposition of a black hole and a thin disk in a magnetic field}

We considerer the superposition of a  Schwarzschild  black hole
($\hat \psi_{S}$, $\hat \Lambda_{S}$)  with a  thin disk ($\hat \psi_{D}$, $
\hat \Lambda_{D}$)  
in presence of a magnetic field. The  Schwarzschild  black hole  metric
functions with mass $m$ are given by 
\begin{subequations}\begin{eqnarray}
\hat  \psi_S & = & \frac {1}{2} \ln \left( {\frac{{x-1}}{{x + 1}}} \right), 
\label{eq:psh} \\
\hat \Lambda_S  & = &  \frac{1}{2} \ln \left[ \frac{x^2-1}{x^2-y^2}  \right ],
\label{eq:Lsh} 
\end{eqnarray}\end{subequations}
where  ($x$, $y$) are prolate spheroidal coordinates
which are related to Weyl coordinates ($\rho$, $z$) by
\begin{subequations}\begin{eqnarray}
\rho ^2 &=& k^2 (x^2  - 1)(1 - y^2 ), \quad  z = k xy,   
\label{eq:prola1} \\
2kx &=&  r_+ + r_-, \quad  2 k y =  r_+ - r_-, \quad  r_{\pm}^2 = \rho^2 + (z \pm
k)^2, \label{eq:prola2}
\end{eqnarray}\end{subequations}
with  $x \geq 1$ and   $-1 <y<1$. For the  black hole $k=m$.
Since  the metric function $\hat \psi$ 
satisfies the   Laplace's equation  and it is linear, then the
superposition   $\hat \psi = \hat \psi_{S} + \hat \psi_{D}$ is also solution.
The other  metric function $\hat\Lambda$ is nonlinear but holds the relation
  \cite{LL2}
\begin{equation}
\hat \Lambda = \hat \Lambda_S + \hat \Lambda_{D} + \hat \Lambda_{SD}, \label{eq:Lambdasuper}
\end{equation}
where
\begin{equation}
\hat  \Lambda_{SD} = 2 \int \rho \{ (\hat \psi_{S,\rho} \hat \psi_{D,\rho}  
- \hat \psi_{S,z}\hat \psi_{D,z} ) d\rho 
+ (\hat \psi_{S,\rho} \hat \psi_{D,z} + \hat \psi_{S,z} \hat \psi_{D,\rho} )dz 
\}.  \label{eq:LSD}
\end{equation}

The magnetic field can be included using
the Ernst's method. The  application of such  procedure to the  system 
(\ref{eq:e-m1})-(\ref{eq:e-m4}) yield 
\cite{GG-CRIS}
\begin{subequations}\begin{eqnarray}
 f &=& \frac{4}{{\left[ {\left( {\beta +1 } \right)e^{ - \hat  \psi }  
- \left( {\beta - 1} \right)e^{\hat  \psi } } \right]^2 }}, \label{eq:f} \\
\Lambda &=& \hat \Lambda = \int \rho \left \{  (\hat \psi^{2},_{\rho}- \hat
\psi^{2},_{z} )d\rho  
+ 2  \hat \psi,_{\rho} \hat \psi,_{z} dz  \right \} ,    \label{eq:Lambda}   \\ 
 A &=&   b \int   \rho  \left \{ - \hat \psi _{,z} d\rho  + \hat \psi _{,\rho
}dz  \right \}   ,  \label{eq:A}  
\end{eqnarray}\end{subequations}
where    $\beta = \sqrt{1+b^2}$, being  $b$  the parameter that controls the
magnetic field. In absence of magnetic field ($b=0$),  these  solutions   reduce 
to the  Weyl vacuum solutions and accordingly these fields can be called  
magnetized Weyl fields.
Thus,  calling $A_S$  the magnetic potential  associated to  the black hole
and $A_D$  the magnetic potential associated  to the  
disk field,  according to (\ref{eq:A}) the   magnetic potential 
for the composite system is given by   
\begin{equation}
 A = A_S + A_D.  \label{eq:Asuper}
\end{equation}
The metric potential $\Lambda$ is the same that  the seed potential and the 
other metric function $\psi$  obtains using  (\ref{eq:f}).

To obtain the components of energy-momentum tensor of the combined system 
 we shall compute  the nonzero
components of $b_{ab}$. From (\ref{eq:b})  we have 
\begin{subequations}\begin{eqnarray}
b_{tt}  &=&  [g_{tt,z}] =  -2 e^{2 \psi} [\psi_{,z} ],  \\
b_{\varphi \varphi } &=&  [g_{\varphi\varphi, z}]  
=  -2 \rho^2  e^{-2 \psi} [\psi_{,z} ], \\
b_{\rho \rho } &=&  b_{z z } = [g_{zz,z}]  
=2 e^{2(\Lambda -\Psi)}([\Lambda{,z}]-[\psi_{,z}]) .
\end{eqnarray}\end{subequations}

From (\ref{eq:f})  $\psi_{,z}=  F \hat \psi_{,z} $ where
\begin{equation}
F = \left[ \frac { (\beta + 1){\rm e}^{- \hat \psi} + (\beta -1) {\rm e}^{\hat \psi } }
{ (\beta + 1){\rm e}^{-\hat \psi} - (\beta - 1) {\rm e}^{\hat \psi } }  \right ].
\end{equation}
Furthermore,  for the black hole we have  $[\hat \psi_{S,z}]=[\Lambda_{S,z}]=0$ whereas   
for the disk    $[\hat \psi_{D,z}]= 2 \hat \psi_{D,z}$,
and $[\Lambda_{D,z}]= 2 \Lambda_{D,z}$. Thus 
\begin{equation}  
[\psi_{,z} ] = [ F    \hat \psi_{,z}] =  F [\hat  \psi_{,z}] =  F \{
[ \hat  \psi_{S,z}] + [    \hat \psi_{D,z}] \} =  2 F \hat \psi_{D,z}. \label{eq:jumppsiz}
\end{equation}

Using  (\ref{eq:Lambdasuper})  and (\ref{eq:Lambda})  we have
\begin{eqnarray}
[\Lambda_{,z} ] & = & [\Lambda_{S,z} ] + [\Lambda_{D,z} ] + [\Lambda_{SD,z} ]   
\nonumber   \\
&=& 2 \Lambda_{D,z} +  [\Lambda_{SD,z} ]   \nonumber   \\
&=& 4  \rho \hat \psi_{D,\rho}  \hat \psi_{D,z} +  [\Lambda_{SD,z} ].     \
\end{eqnarray}
But  from  (\ref{eq:LSD})
$\Lambda_{SD,z} = 2 \rho ( \hat \psi_{S,\rho} \hat \psi_{D,z} + \hat \psi_{S,z}
\hat \psi_{D,\rho})
$, then 
\begin{equation}
[\Lambda_{SD,z}] = 2 \rho \hat \psi_{S,\rho} [\hat \psi_{D,z}]= 4 \rho \hat 
\psi_{S,\rho}
\hat \psi_{D,z},
\end{equation}
and hence  
\begin{equation}
[\Lambda_{,z} ] = 4 \rho (\hat \psi_{S} + \hat \psi_{D} )_{,\rho} \hat
\psi_{D,z}.
\label{eq:jumpLz}
\end{equation}
 Using (\ref{eq:jumppsiz})  and (\ref{eq:jumpLz}) we have 
\begin{subequations}\begin{eqnarray}
b_{tt}  &=&  - 4  e^{2 \psi} F \hat \psi_{D,z} ,  \label{eq:btt}  \\
b_{\varphi \varphi } &=&  - 4 \rho^2  e^{-2 \psi} F \hat \psi_{D,z} ,  \label{eq:bvv}    \\
b_{\rho \rho } &=& 4 e^{2(\Lambda - \psi)} \{ 2\rho (\hat \psi_{S} + \hat
\psi_{D})_{,\rho}  -F \} \hat \psi_{D,z}.    \label{eq:bzz}
\end{eqnarray}\end{subequations}

From  (\ref{eq:Qab}), (\ref{eq:Sab}), (\ref{eq:enpr}) and 
(\ref{eq:btt})-(\ref{eq:bzz}), we obtain 
\begin{eqnarray}
\epsilon & =& - S^t_t = \frac{1}{2 \pi} e^{ \psi -\Lambda } (F - \rho (\hat \psi_{S} + \hat \psi_{D} )_{,\rho} )
\hat \psi_{D,z} , \label{eq:ener} \\
p_{\varphi} &=& S^{\varphi}_{\varphi} = \frac{1}{2 \pi} e^{ \psi -\Lambda } \rho (\hat \psi_{S}
+ \hat \psi_{D} )_{,\rho}   \hat \psi_{D,z},  \label{eq:pr}
\end{eqnarray}
where all the quantities are evaluated at $z=0^+$.

Similarly, for the magnetic potential   we have $ [A_{,z}] = [A_{S,z}] + [A_{D,z}]$,
but  from (\ref{eq:A})    $[A_{S,z}] = [b \rho \hat \psi _{S, \rho}] =0$,
so that    $ [A_{,z}] = [A_{D,z}]$.  Thus,  from (\ref{eq:ja}),  (\ref{eq:jatrue}) and  (\ref{eq:j})  
one finds that  if $A_{D,z}$ is assumed  discontinuous  across  the disk  the current
density  is  given by  
\begin{equation}
 \text{\sl j}  =  - \frac{1}{2 \pi \rho}  e^{2 \psi -\Lambda}  A_{D,z} .
\label{eq:corr}
\end{equation}

The speed $v$ of counterrotating of the  particles in the disk is given by 
\begin{equation}
 v^2 = \frac {F \rho (\hat \psi_{S} + \hat \psi_{D} )_{,\rho}} { 1 - F \rho (\hat \psi_{S} + \hat \psi_{D} )_{,\rho} 
 + \frac{1}{8 \pi} b^2 f }.   \label{eq:v2super}
\end{equation}


\section{Some simple examples}

\subsection{Black hole surrounded by Kuzmin-Chazy-Curzon  disks in a magnetic
field}

Exact solutions which represent the field of a  disk 
can be obtained using  the well known ``displace,
cut and reflect'' method  that was first
used by Kuzmin \cite{KUZMIN} and Toomre \cite{TOOMRE} to constructed Newtonian
models of disks, and later extended to general relativity 
\cite{BLK,BLP,BL,GL2}. Given a solution of the
Einstein-Maxwell equation, this procedure is mathematically equivalent to apply
the transformation $z \rightarrow |z| + z_0$,
with $z_0$ constant. The  resulting disks are essentially 
of infinite extension.  This method applied to the Chazy-Curzon  solution \cite{CH,C}
produces a  Kuzmin-Chazy-Curzon disk  of  mass $M$ with  metric functions  
\begin{subequations} \begin{eqnarray}
 \hat \psi _{D} &=& - \frac{M}{\sqrt{ \rho^2 + (|z| + z_0)^2  } },  \\
\hat \Lambda _D &=& -\frac{M ^2 \rho^2 }{2 [ \rho^2 + (|z| + z_0)^2 ]^2}  .
 \end{eqnarray}\end{subequations}
This solution  is the relativistic generalization of the  newtonian Kuzmin  disk.

Now we consider the superposition of a black hole and this disk in a magnetic field. 
For this system the magnetic potential (\ref{eq:Asuper}) is 
\begin{equation}
A =  b  m (y + 1)  +  bM( \frac{|z| + z_0}{ \sqrt{ \rho^2 + (|z| + z_0)^2 }  }   + 1 ),
\end{equation}
where the  first term  on right-hand side is $A_S$ and the second $A_D$.
The interaction term $\Lambda_{SD}$ (\ref{eq:LSD}) between
the disk and the black hole can be calculated  
in prolate coordinates. In this coordinates the expression  for $\Lambda$ (\ref{eq:Lambda})
takes the form
\begin{subequations}\begin{eqnarray}\label{eq:Afini}
 \Lambda _ {,x}  &=& \left( \frac{ 1 - y^2 }{ x^2  - y^2 } \right) \left[ x (x^2 
- 1)  \hat \psi _{,x} ^2  
- x ( 1 - y^2 ) \hat \psi _{,y} ^2  - 2y ( x^2  - 1 ) \hat \psi _{,x} \hat \psi
_{,y } \right] ,   \label{eq:Lambdax} \\
\Lambda _ {,y}  &=& \left(  \frac{x^2  - 1}{x^2  - y^2 } \right ) \left[ y ( x^2
 - 1 ) \hat \psi _{,x} ^2 
- y ( 1 - y^2 ) \hat \psi _ {,y} ^2  + 2x ( 1 - y^2 ) \hat \psi _{,x} \hat \psi
_{,y}   \right],     \label{eq:Lambday}
\end{eqnarray}\end{subequations}
and thus  from (\ref{eq:Lambday}) we obtain
  \begin{equation}
\Lambda_{SD} = 2 \int _{-1}^y \left(  \frac{x^2  - 1}{x^2  - y^2 } \right ) 
\left \{ 
[ y ( x^2 - 1 ) \hat \psi _{D,x}  + x ( 1 - y^2 ) \hat \psi _{D,y} ] \hat \psi
_{S,x}
+ (1-y^2)(x \hat \psi _{D,x} -y \hat \psi _{D,y} ) \hat \psi _{S,y}
 \right \} dy ,  
\end{equation}
where the integral limits are chosen by requiring that the function $ \Lambda $
to be regular on the axis of symmetry. But  $\hat \psi_{S,y}=0$ and $\hat
\psi_{S,x}=1/(x^2-1)$, then
\begin{equation}
\Lambda_{SD} =  2 \int _{-1}^y \frac{1}{(x^2-y^2)} 
\left [  y ( x^2 - 1 ) \hat \psi _{D,x}  + x ( 1 - y^2 )  \hat \psi _{D,y} 
\right ] dy.   \label{eq:LSDprola}
\end{equation}

For $z>0$ this integral gives
\begin{equation}
\Lambda_{SD} =  \frac{2  M m }{ [z_0^2-m^2] \sqrt{ \rho
^2 + (z + z_0)^2 }  }  \left [ mx + z_0y
- \sqrt{ \rho ^2 + (z + z_0)^2 } \right ],  \label{eq:LSDcurzon}
\end{equation} 
or  in Weyl coordinates 
\begin{eqnarray}
\Lambda_{SD} &=&  \frac{M}{ [z_0^2-m^2] \sqrt{ \rho ^2 + (z +
z_0)^2 }  }  
\left \{   [z_0 + m]\sqrt{ \rho ^2 + [z + m]^2 }  \right .
\nonumber  \\
 & & \left .   - [z_0 - m]\sqrt{ \rho ^2 + [z - m]^2 }
- 2m \sqrt{ \rho ^2 + (z + z_0)^2 } \right \} .
\end{eqnarray}

From (\ref{eq:ener})-(\ref{eq:v2super}), the main  physical quantities associated
with the system  are
\begin{eqnarray}
\tilde \epsilon & =&  \frac{ 4 M z_0 e^{ \psi -\Lambda } } { (\rho^2 +
z_0^2)^{3/2} } \left [F -   \frac{m }
{ \sqrt {\rho^2 + m^2} }  - 
 \frac{M \rho^2 }{ (\rho^2 + z_0^2)^{3/2}  } \right ] , \\
\tilde p_{\varphi} &=& \frac{ 4 M z_0 e^{ \psi -\Lambda } } { (\rho^2 +
z_0^2)^{3/2} } \left [   \frac{ m }
{ \sqrt {\rho^2 + m^2} }  + 
 \frac{M \rho^2 }{ (\rho^2 + z_0^2)^{3/2}  }     \right ] , 
\label{eq:prcur} \\
\text{\sl j}& = &  - \frac{M b}{2 \pi}  e^{2\psi -\Lambda}  \frac{\rho
}{(\rho^2 + z_0^2)^{3/2}} , \label{eq:corrcur}  \\
v^2 & = &  \frac{F  \left [   \frac{ m }{ \sqrt {\rho^2 + m^2} }  + 
 \frac{M \rho^2 }{ (\rho^2 + z_0^2)^{3/2}  }     \right ] }
 { 1- F\left [   \frac{ m }{ \sqrt {\rho^2 + m^2} }  + 
 \frac{M \rho^2 }{ (\rho^2 + z_0^2)^{3/2}  }     \right ] 
 +  \frac{1}{8 \pi} b^2 f } ,  \label{eq:v2-curzon}
\end{eqnarray}
with $ \tilde \epsilon = 8 \pi \epsilon $ and  $ \tilde p = 8 \pi p $.

\begin{figure}
$$
\begin{array}{ccc}
\tilde \epsilon    & \tilde  p_\varphi &  -\text{\sl j} \\
\includegraphics[width=0.3\textwidth]{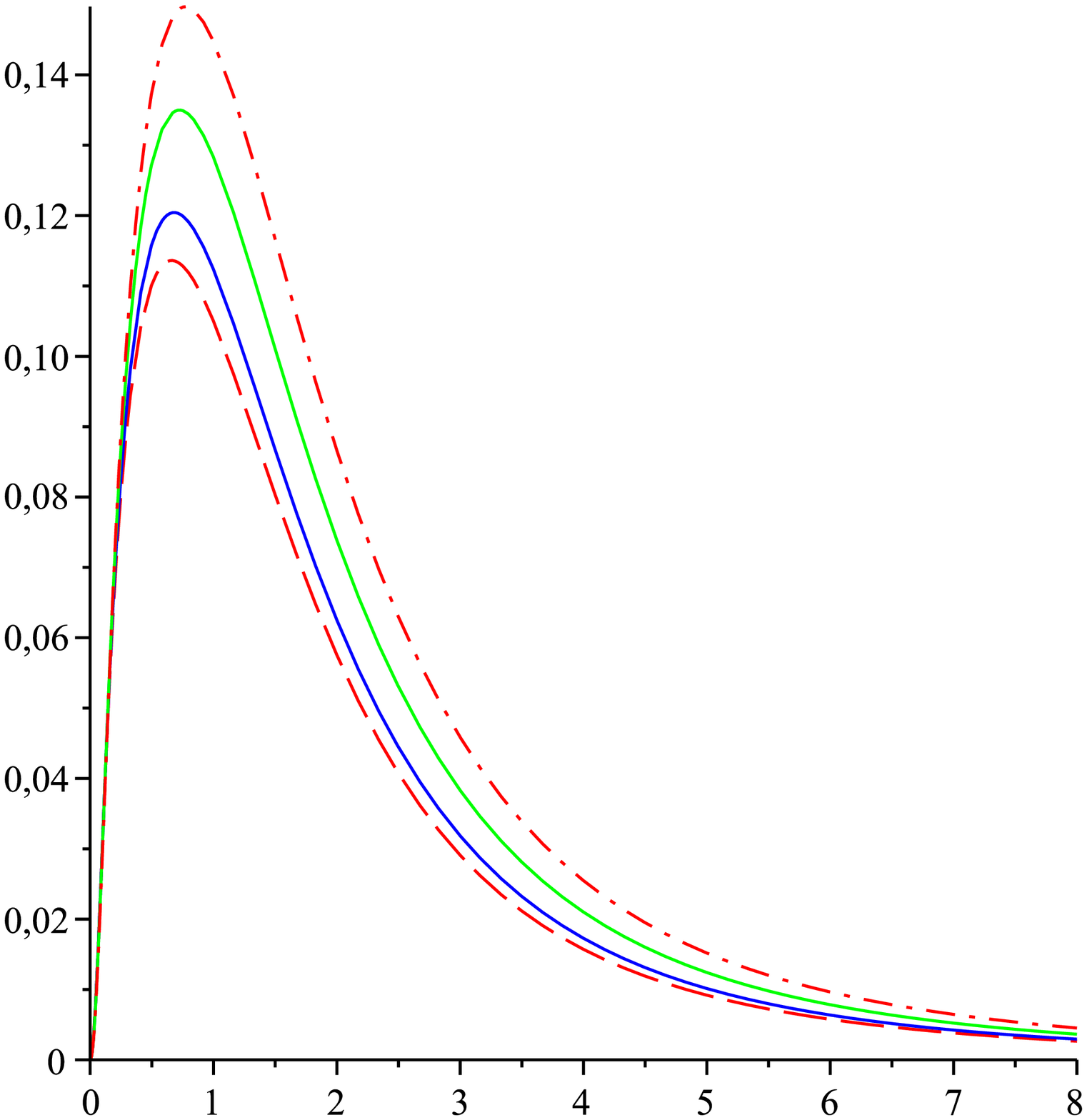} &      
\includegraphics[width=0.3\textwidth]{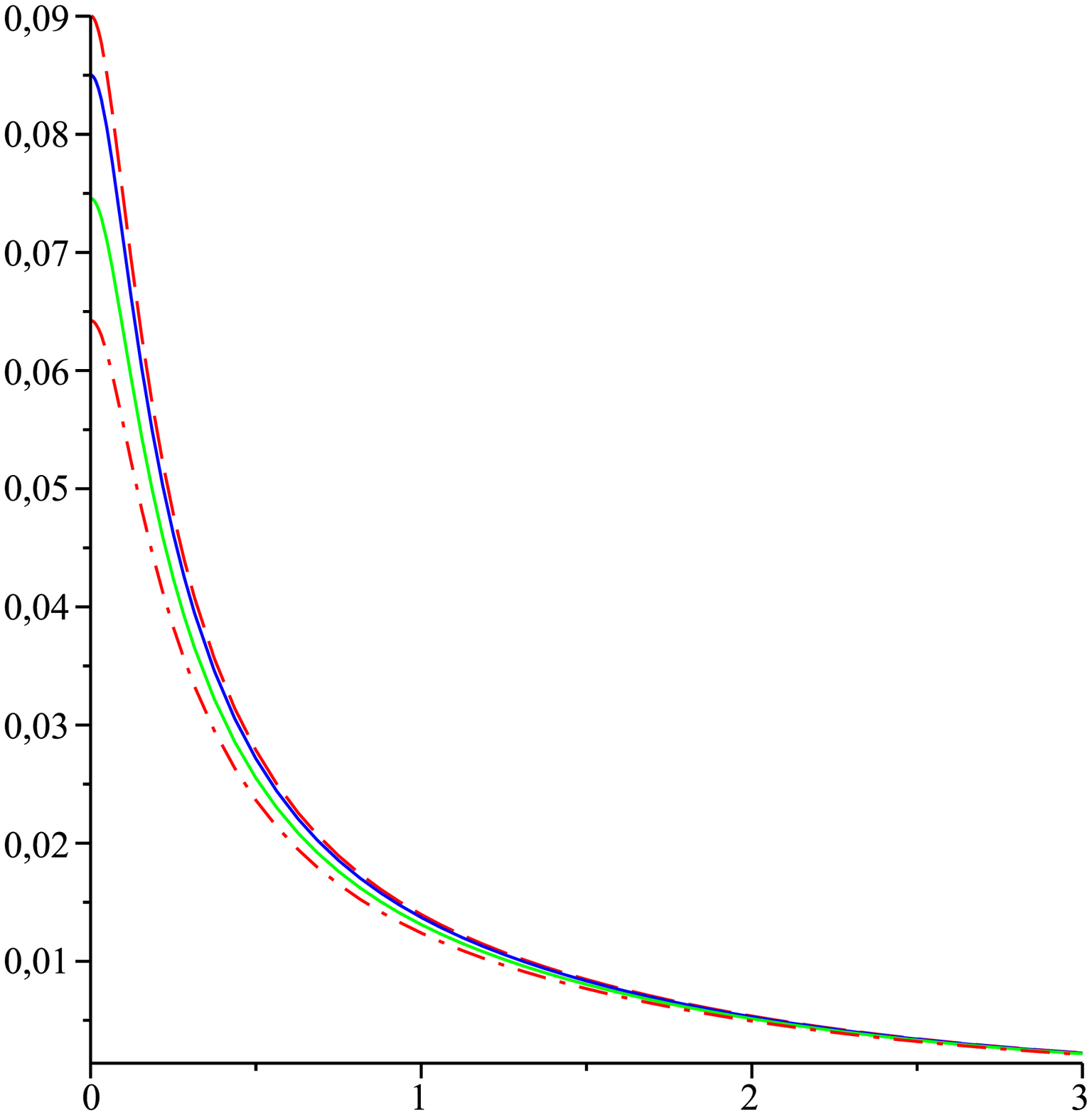} &
\includegraphics[width=0.3\textwidth]{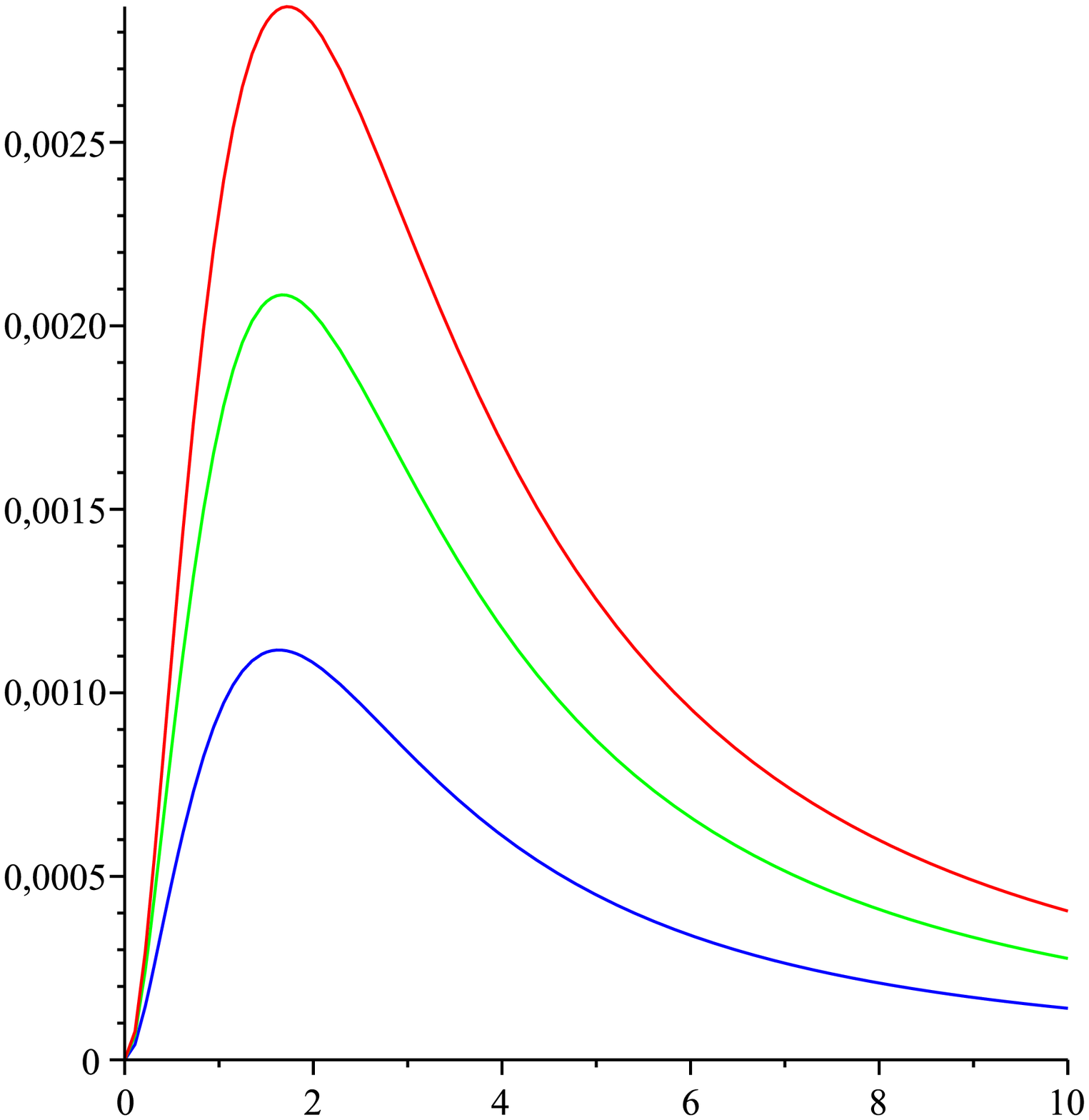}   \\ 
(a)    &  (b) & (c) 
\end{array}
$$	
\caption{$(a)$  The  energy density $\tilde \epsilon$ and   $(b)$ the azimuthal
pressure $\tilde p_\varphi$ for the system  black hole 
and a  Kuzmin-Chazy-Curzon  infinite disk in a magnetic field  
with $z_0=2$,  $m =0.1$,  $M = 0.2$,  and
for  values of magnetic field parameter  $b= 0$ (dashed curves), $0.5$,  $1$,
and $1.5$ (dash-dotted  curves), as  functions of $\rho$. $(c)$ The azimuthal electric
current density $\text{\sl j}$ for
$b= 0$ (axis $\rho$), $0.5$,  $1$, and $1.5$ (top curve) and the same value of
other parameters.}
\label{fig:curzon1}
\end{figure}

\begin{figure}
$$
\begin{array}{cc}
v^2   &  h^2  \\
\includegraphics[width=0.3\textwidth]{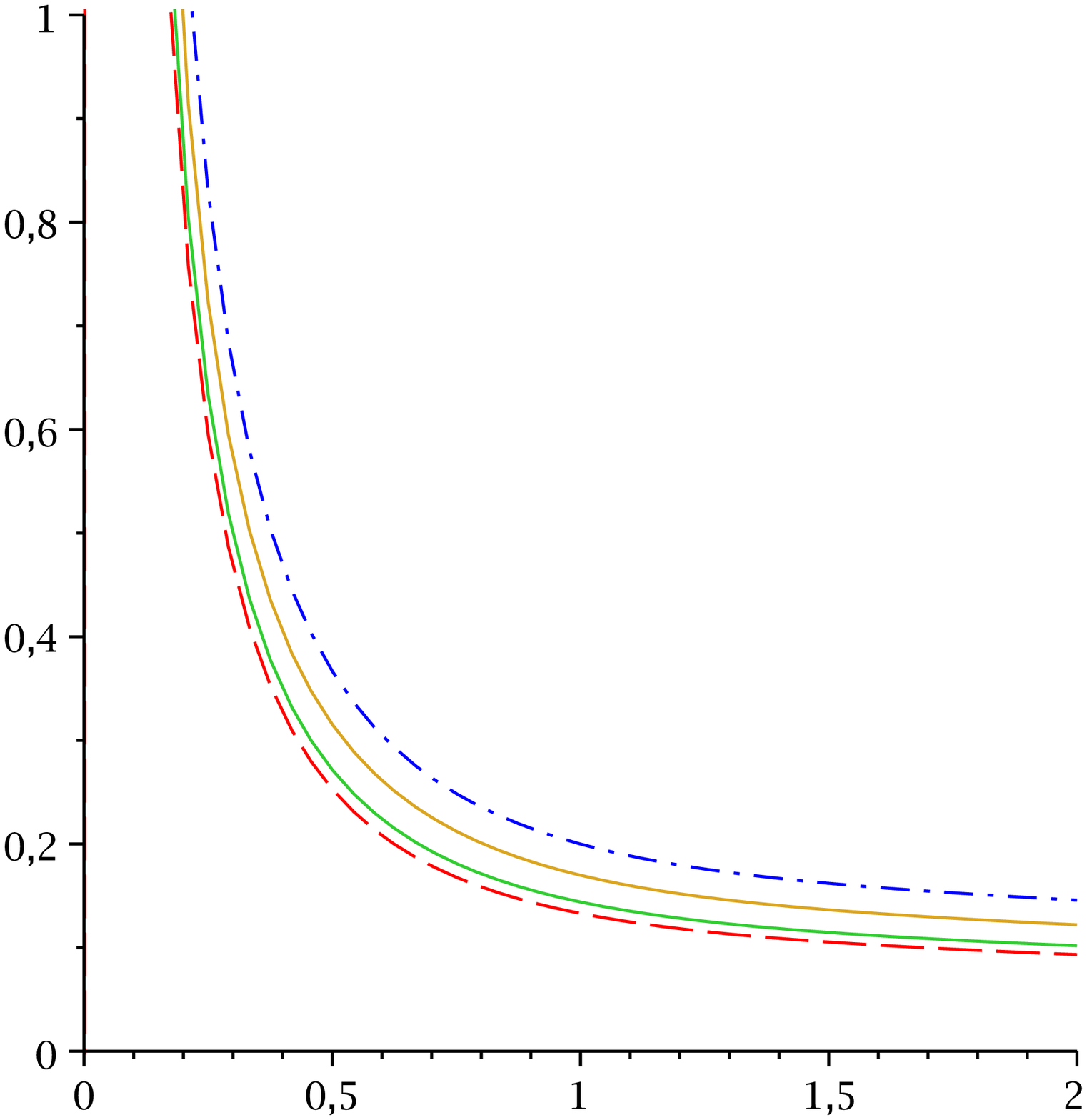} &      
\includegraphics[width=0.3\textwidth]{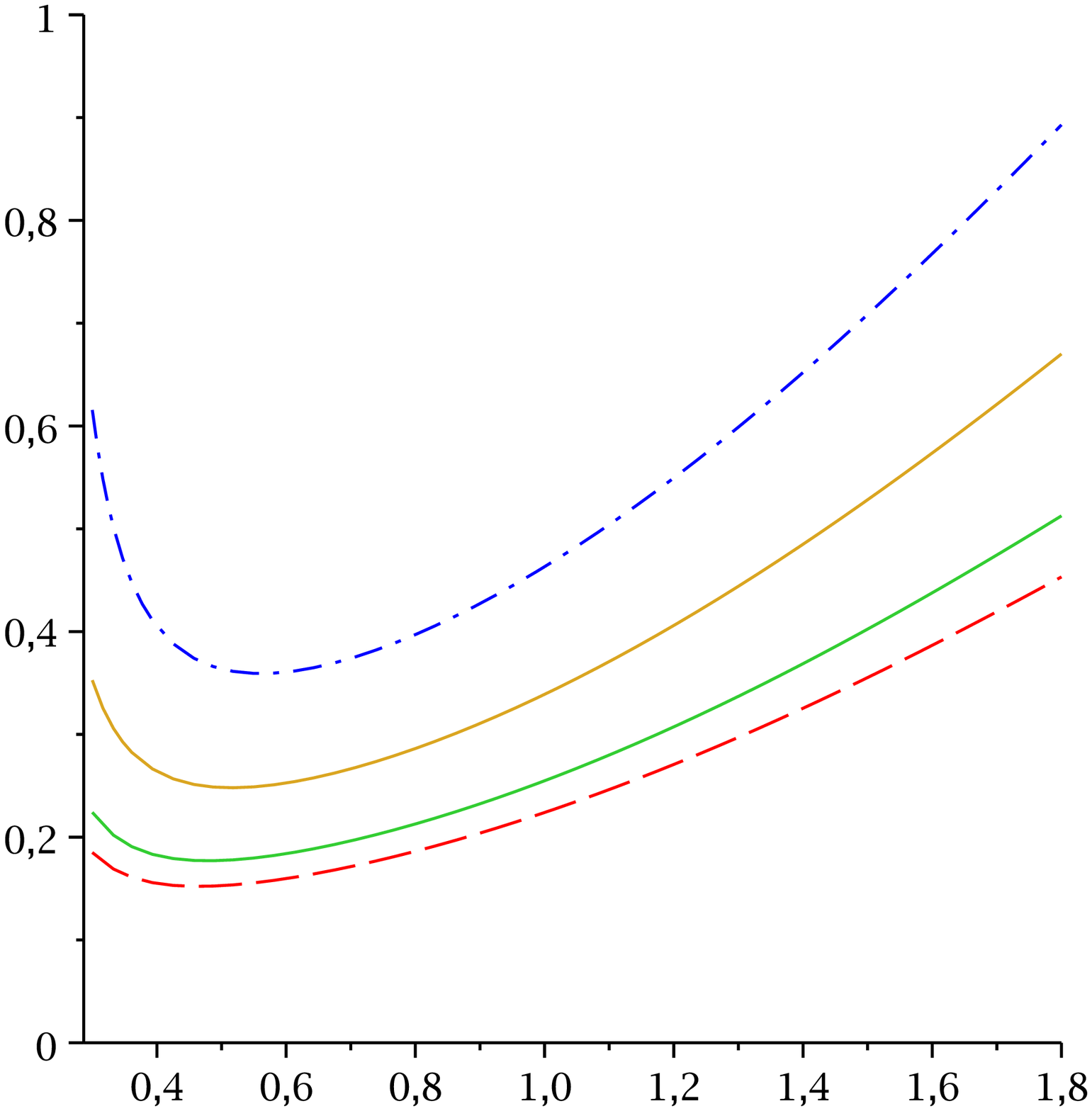}  \\ 
(a)    &  (b)  
\end{array}
$$	
\caption{$(a)$  The speed   $v^2$ of counterrrotation and
$(d)$ the specific angular momentum $h^2$   
for the system  black hole and a  Kuzmin-Chazy-Curzon  infinite disk in a
magnetic field  
for   $b= 0$ (dash curves), $0.5$,  $1$, and $1.5$ (dash-dotted  curves) and the same
value of other parameters.}
\label{fig:curzon2}
\end{figure}

Figs.  \ref{fig:curzon1} and  \ref{fig:curzon2} show curves of 
the  energy density $\tilde \epsilon$, the azimuthal
pressure $\tilde p_\varphi$,  the azimuthal electric current density $\text{\sl j}$,
the speed   $v^2$ of counterrrotation and
the specific angular momentum $h^2$ as  function of
$\rho$    for  $z_0=2$,  $m =0.1$,  $M = 0.2$,  and different 
values of magnetic field parameter  $b= 0$, $0.5$,  $1$, and $1.5$.
We  see   that  energy density is a positive quantity  in
concordance with  the weak energy condition, as well as the stress in
azimuthal  direction (pressure). Eq. (\ref{eq:prcur}) shows that we have
pressure  for all values of parameters.  
Since $ \epsilon + p_\varphi >0 $, the strong energy condition is also
satisfied.
These properties characterize a distribution  of matter
with the usual gravitational attractive property. 
However, the dominant energy condition ($v\leq 1$)  is not satisfied  
in the central region
located between the black hole and the photon radius, 
and  the magnetic field enhances the zone  of superluminal  speed. 
Since   the function $F$ tends to one  and $f$ to zero as   $\rho$ tends a zero,   
the expression (\ref{eq:v2-curzon}) shows that these   systems present  the same  behavior 
for all values of the parameters.
We also observer that these structures  present  strong instabilities  inside
the photonic orbit, and that  the magnetic field  increases the region of instability.

\subsection{BLACK HOLES SURROUNDED BY FINITE DISKS  IN AN MAGNETIC FIELD}

Solutions representing the field of a finite thin disk can be obtained resolving
the Laplace equation in oblate  spheroidal coordinates
($u$,$v$), which are related to prolate
coordinates by $x=-iu$,  $y=v$, and  $k=i a$, where $a$ is the radius of the
disk, 
and with the   Weyl coordinates ($\rho$, $z$) by 
\begin{subequations}\begin{eqnarray}
 \rho ^2 & =& a ^2 (u^2  + 1)(1 - v^2 ) ,  \\
 z & = & auv, 
\end{eqnarray}\end{subequations}
with $u \geq 0$ and  $-1< v <1$. 
These  solutions are known in the literature as the Morgan-Morgan solutions \cite{MM1}  
and the metric function $\hat \psi_D$ is given by 
\begin{equation}
\hat  \psi_D  = - \sum\limits_{n = 0}^\infty  {c_{2n} q_{2n} (u)P_{2n} (v)} , 
\label{EQ:serieobla}
\end{equation}
where  $c_{2n}$ are constants, $P_{2n}$ are the Legendre polynomials of order $2n$ and
\begin{equation}
q_n (u) = i^{n + 1} Q_n (iu),
\end{equation}
being  $Q_n (iu)$  the Legendre functions of the second kind.
For $n=0$ we have  the zeroth order Morgan-Morgan
disk and for the terms $n = 0$ and $n = 1$
the first Morgan-Morgan disk \cite{LETSTAB}.
In this case,  the metric functions  are
\begin{subequations}\begin{eqnarray}
\hat  \psi_D   &=& - (M/a) \cot^{ - 1} (u) - \frac{1}{4} (M/a)  
\left( 3v^2  - 1 \right) \left[ (3u^2  + 1) \cot^{ - 1} (u) - 3u \right], \\
\hat \Lambda_D  &=& 
-\frac{9 M^2}{16 a^4} \rho^2 (9u^2v^2+v^2 -u^2 -1)  
\nonumber \\
&& +\frac{9 M^2}{8 a^2 }u(1-v^2)(9u^2v^2 + 7v^2 - u^2 +1) \cot^{ - 1}(u)  \nonumber \\
&& - \frac{9 M^2}{16 a^2 }(1-v^2)(9u^2v^2 + 4v^2 -u^2 + 4),   
\end{eqnarray}\end{subequations}
where  $M$ is the mass  of the disk.

We consider the   system composed  of a black hole and a disk in a magnetic field.
The magnetic potential is given by 
\begin{equation}
 A  =   b  m (y + 1) +  a b (M/a) \left( v + 1 \right) + \frac{1}{2} a b (M/a) v\left( 1 - v^2 
\right)\left[  3u\left( u^2  + 1 \right)\cot^{ - 1} (u) - 3u^2  - 2 \right],
\label{eq:Afini}
\end{equation}
and, in oblate coordinates, the interaction term (\ref{eq:LSDprola})    takes the form
\begin{equation}
\Lambda_{SD} =  \frac{2 k}{a} \int _{-1}^v    
 \left (  \frac{ v ( u^2 + 1 ) \hat \psi _{D,u}  + u ( 1 - v^2 )  \hat \psi
_{D,v}} {u^2  + v^2}   \right ) dv. 
\end{equation}
For $z>0$  this integral gives
\begin{equation}
\Lambda_{SD} = \frac{ 3 m M}{a^2} (1-v^2) \left (  u \cot^{
-1}(u) - 1  \right ).
\end{equation}

In oblate coordinates the disk is located in  $u=0$,  $-1< v <1$, 
and   when the disk is crossed the coordinate $v$ changes of sign but not its
value absolute, whereas $u$ is continuous. 
Hence   the gravitational field is continuous across the disk
whereas  its normal derivate is discontinuous. 
The same does not occur with the magnetic potential
which  is discontinuous across the disk
but its first derivate is continuous in the direction normal to 
the disk. This implies that   the source of the magnetic field   
is non planar but of a different origin such as a remnants or
fossil magnetic field \cite{FOSSIL}, or can come from
external sources, such as the presence of a nearby magnetars or 
neutron stars.  
These models   can be interpreted as  the superposition of a black hole  
and  a finite disk  immersed in a magnetic field (see Appendix). 
From (\ref{eq:ener})-(\ref{eq:corr})  we obtain  
\begin{eqnarray}
\tilde \epsilon & =& \frac{12 M}{  a^3} e^{ \psi -\Lambda } \sqrt{1-
\frac{\rho^2}{a^2}}  
\left [ F -   \frac{3 \pi}{4} \frac{M \rho^2}{ a^2} - \frac{m }{
\sqrt{m^2 + \rho^2}}    \right ] ,   \label{eq:enfini} \\
\tilde p_{\varphi} &=& \frac{12 M}{ a^3} e^{ \psi -\Lambda }  \sqrt{1-
\frac{\rho^2}{a^2}}   
\left[   \frac{3 \pi}{4} \frac{M \rho^2}{ a^2} + \frac{m}{
\sqrt{m^2 + \rho^2}}     \right ] ,   \label{eq:prfini} \\
v^2 & = & \frac { F  \left [  (3 \pi / 4) M \rho^2 / a^2  +  m /   \sqrt{ 
 m^2  + \rho^2  }   \right ] }
{  1 -  F  \left[ (3 \pi / 4) M \rho^2 /  a^2  + m /  \sqrt{ 
m^2  + \rho^2  }  \right ] }.     \label{eq:v2fini}
\end{eqnarray}

Now, in order to construct a planar source for the magnetic field is necessary   
to include the electric field. The electric potential is given by 
\begin{equation}
\phi = c  \left[ \frac { {\rm e}^{- \hat \psi} +  {\rm e}^{\hat \psi } }
{ (\beta + 1){\rm e}^{-\hat \psi} - (\beta - 1) {\rm e}^{\hat \psi } }  \right ], 
\end{equation}
where $\beta ^2 = 1 + b^2 + c^2$, being $c$ the parameter that controls the  electric field.
The  energy density and the pressure have the same form that the expressions (\ref{eq:enfini}) and (\ref{eq:prfini})  
and   the only non-zero component of the current density 
$\mbox{\sl j}_a$ on the plane $z=0$ is 
\begin{equation}
\mbox{\sl j}_t = \ - \frac{1}{2 \pi} e^{\psi - \Lambda} \phi _{,z}. 
\label{eq:corelec}  
\end{equation}
In terms of the tetrad (\ref{eq:tetrad1}) - (\ref{eq:tetrad2}) the  electric charge density 
$\sigma$   is given  by
\begin{equation}
\sigma = -V^0 \mbox{\sl j}_t.
\label{eq:sigma}
\end{equation} 

However, since there is no electric current even  generating the magnetic field,  we need 
assume the counterrotating hypothesis. Using (\ref{eq:Sabsum}) - (\ref{eq:jacon}) and (\ref{eq:elecgeo})
we calculate  the physical quantities associated to the  two streams
\begin{subequations}\begin{eqnarray}
(u^0_\pm)^2 \epsilon_\pm & = & \frac{g^{00} \omega_\pm \epsilon}{(\omega_\pm -  \omega _\mp)}, \\
u^0_\pm  \sigma_\pm  & = & - \frac{V^0 \omega_\mp \sigma}{(\omega_\pm -  \omega _\mp)},  \\
\omega _ \pm & = & \frac{ -T_2 \pm  \sqrt{T_2^2 - T_1 T_3 } } { T_1 } ,
\end{eqnarray}\end{subequations}
where
\begin{subequations}\begin{eqnarray}
T_1  &=&  g_{\varphi \varphi, \rho} ,  \\
T_2 & = & - \frac{ \text{\sl j}_0  A_{, \rho} }{\epsilon} ,  \\
T_3 & = & g_{tt, \rho} -    \frac{ 2  \text{\sl j}_0  \phi_{, \rho} }{\epsilon},
\end{eqnarray}\end{subequations}
and $u^0_\pm$ is given by (\ref{eq:u0}). The counterrotating speed $v$ can be obtained from
(\ref{eq:vc2}).

\begin{figure}
$$
\begin{array}{cc}
\tilde \epsilon    & \tilde  p_\varphi  \\
\includegraphics[width=0.3\textwidth]{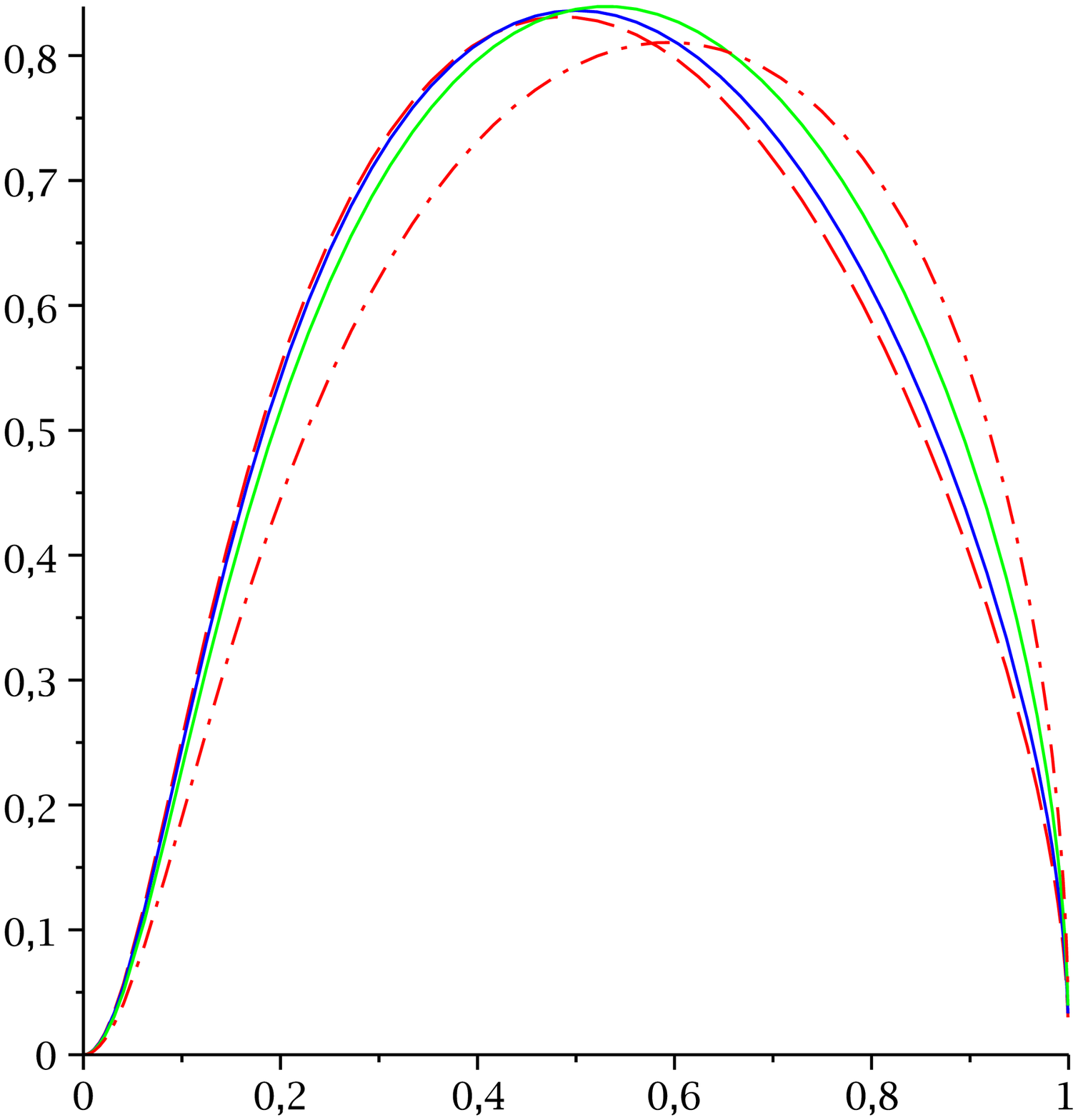} &      
\includegraphics[width=0.3\textwidth]{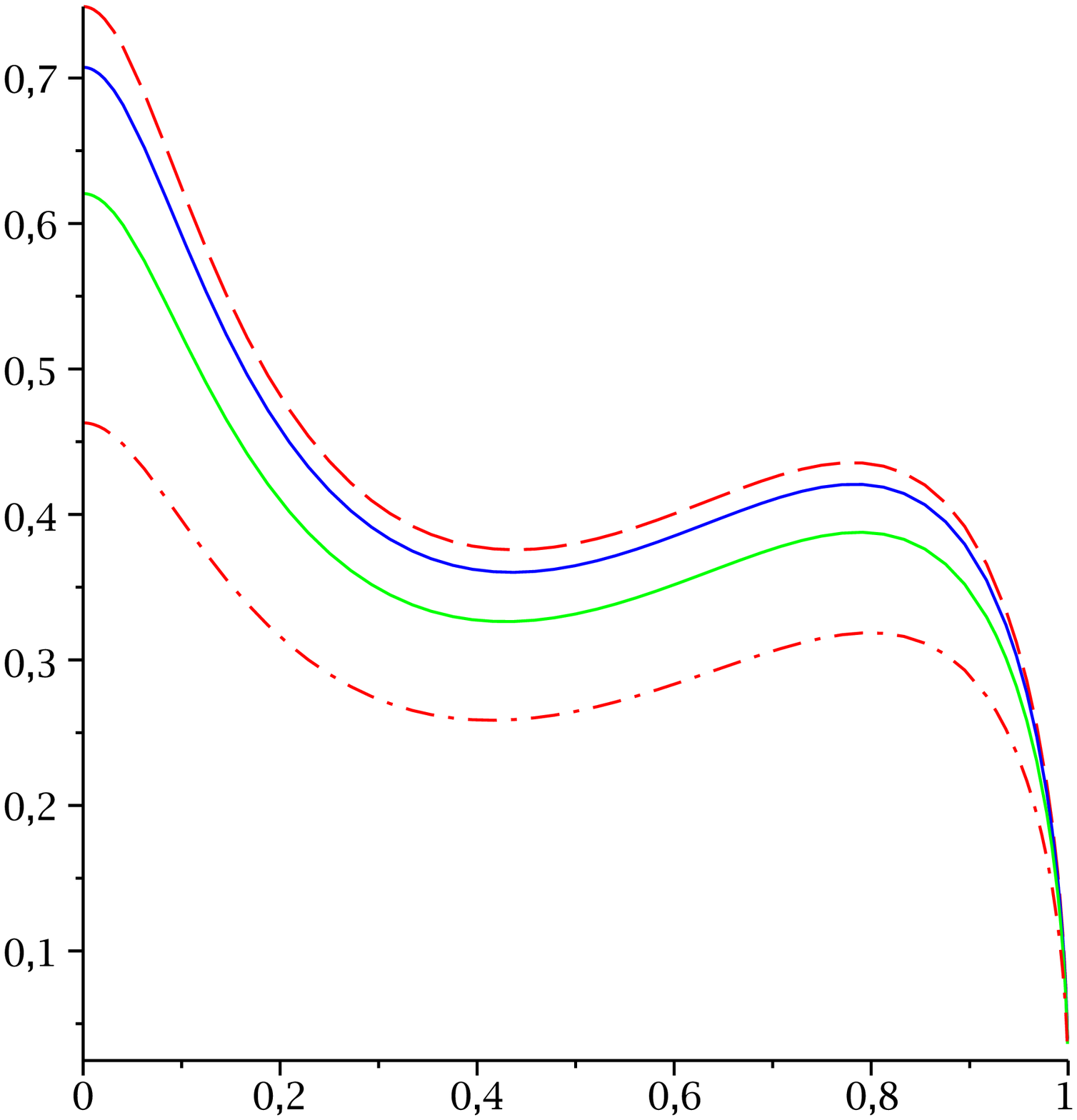}  \\ 
(a)    &  (b)  \\
& \\
v^2 & h^2 \\
\includegraphics[width=0.3\textwidth]{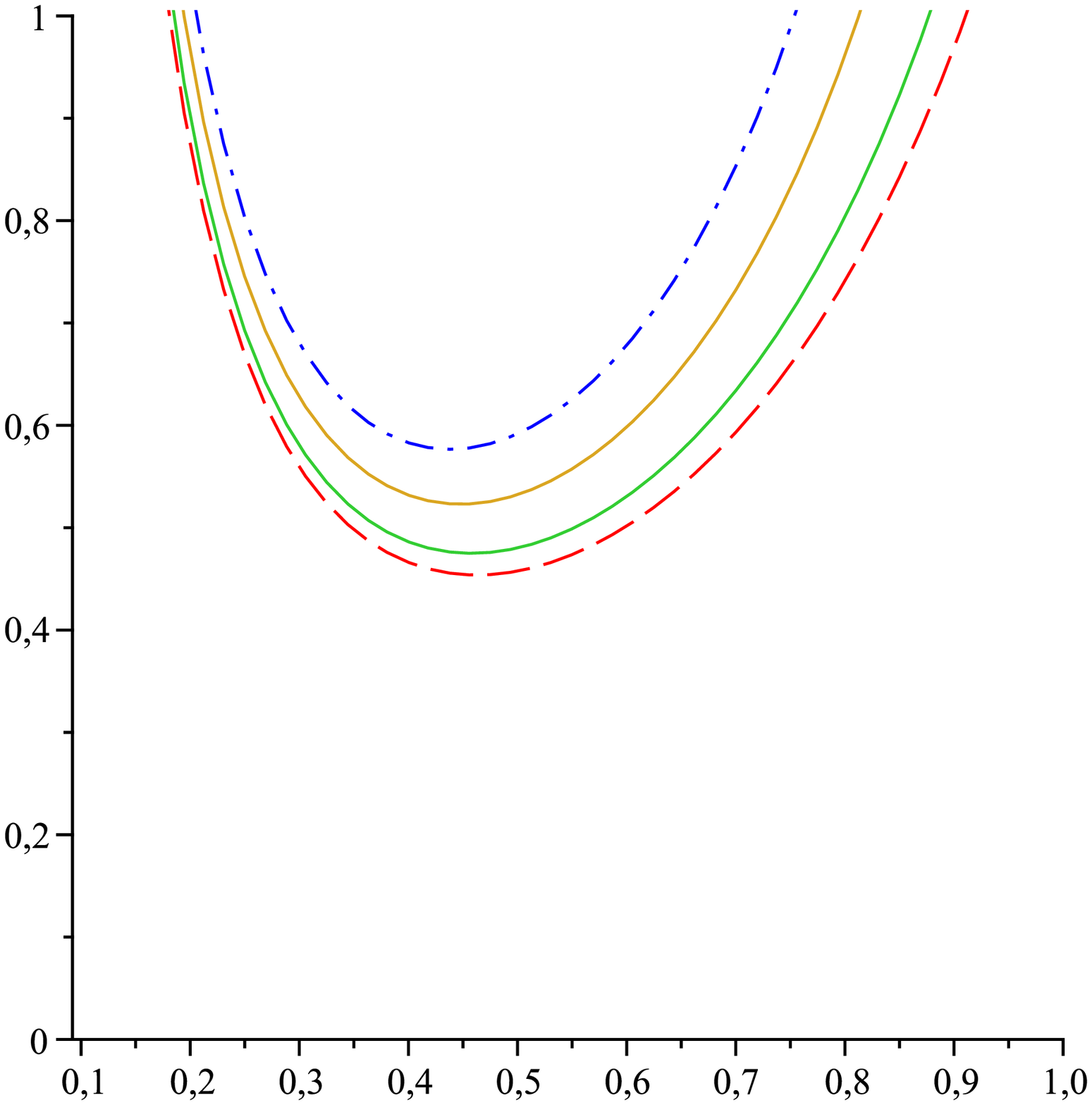} &
\includegraphics[width=0.3\textwidth]{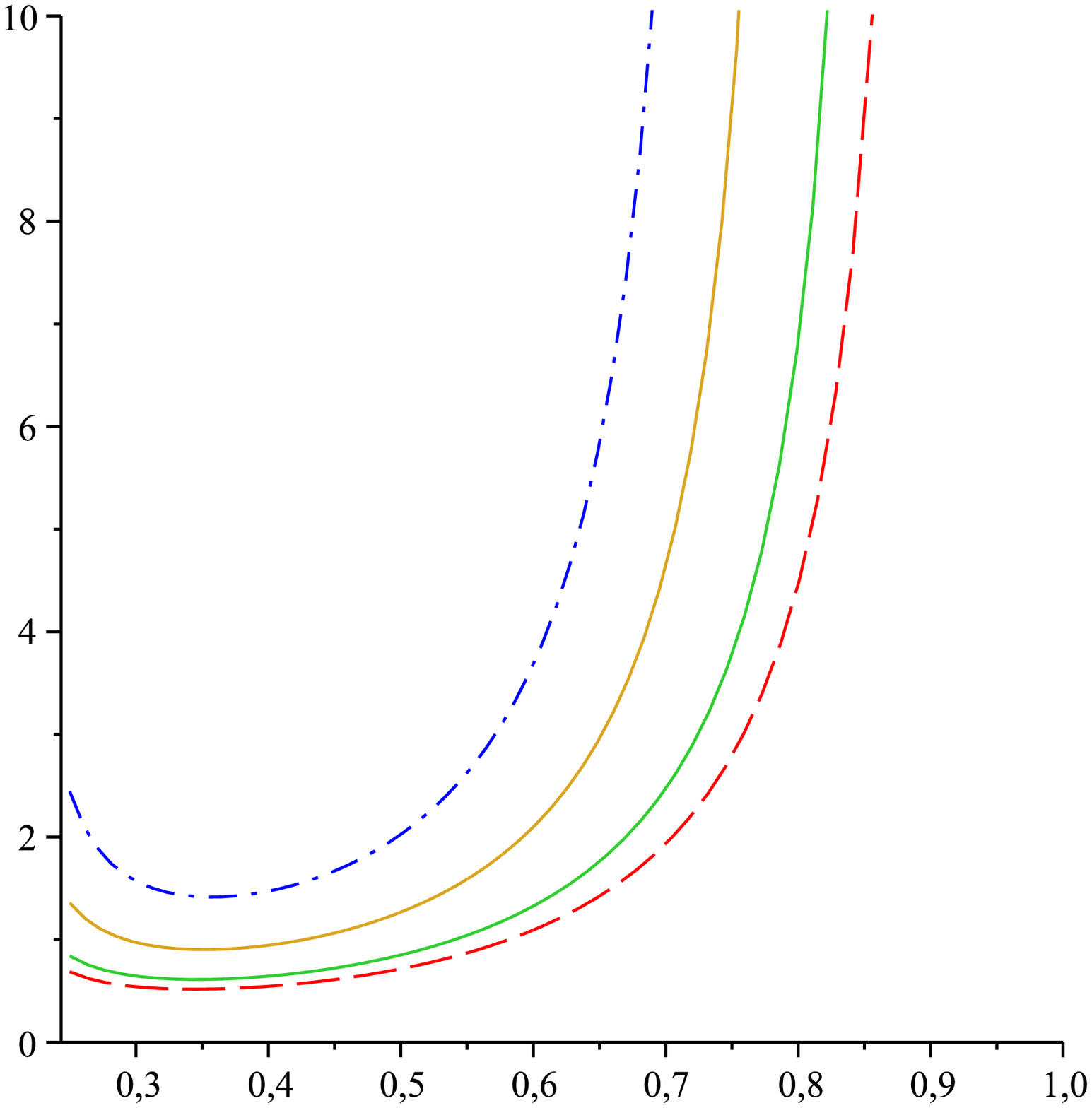}  \\ 
(c)    &  (d) 
\end{array}
$$	
\caption{$(a)$  The  energy density $\tilde \epsilon$,  $(b)$ the azimuthal
pressure $\tilde p_\varphi$,  $(c)$ the speed   $v^2$ of counterrrotation, and
$(d)$ the specific angular momentum $h^2$   
for the system  black hole and the  first  Morgan-Morgan finite  disk in a
magnetic field   with  radius  $a=1$, $m =M=0.1$,  and
for  values of magnetic field parameter  $b= 0$ (dashed curves), $0.5$,  $1$,
and $2$ (dash-dotted  curves), as  functions of $\rho$. }
\label{fig:finitos}
\end{figure}

In Fig. \ref{fig:finitos} we show  the  energy density
$\tilde \epsilon$,  the azimuthal
pressure $\tilde p_\varphi$, the speed   $v^2$ of counterrrotation, and the
specific angular momentum $h^2$  
for a  system consisting of a  black hole and the   first 
Morgan-Morgan  finite  disk in presence of  a magnetic field  with radius  
$a=1$, $m =M=0.1$,  and
different  values of magnetic field parameter  $b= 0$ (dashed curves), $0.5$,  $1$,
and $2$ (dash-dotted  curves), as  functions of $\rho$. 
We find  that the  energy density is a positive quantity  in
agreement  with  the weak energy condition, and we have pressure
for all values of the  parameters. 
However, the dominant energy condition ($v\leq 1$)  is  not satisfied  and the  presence 
magnetic field  enhances the zones of superluminal  speed. 
Such   behavior is also  observed   when the electric field is included. 
We also  see   that  the magnetic field  increases   the regions of
instability.



\subsection{A model of AGN }
We now consider  a simple model of  active galactic nuclei
consisting of black hole, a Kuzmin-Chazy-Curzon disk and two rods 
of linear mass density $\lambda$
located along $[-z_2 , -z_1]$ and $[z_1 , z_2]$ on the $z$ axis, 
in presence of magnetic field. 
The metric potentials of a finite rod  located on the $z$ axis between $z=z_1$
and 
$z=z_2$ ($z_2 > z_1$ ) are  \cite{LV-AGN}
\begin{subequations}\begin{eqnarray}
 \hat \psi_R & = & \lambda \ln  \left ( \frac{\mu _1}{\mu _2} \right ), \\
\hat  \Lambda _R &= & 2 \lambda ^2 \ln \left [  \frac{(\rho^2 + \mu_1 \mu_2)^2
}{ (\rho ^2 + \mu_1^2)(\rho ^2 + \mu_2^2) }
\right ] ,
\end{eqnarray}\end{subequations}
with $\mu _1 = z_1-z + \sqrt{\rho^2 + (z_1-z)^2}$ and  $\mu _2 = z_2-z +
\sqrt{\rho^2 + (z_2-z)^2}$,
and for the  combined system the metric potential   $\hat \psi$  is
\begin{equation}
 \hat \psi =   \hat \psi_s + \lambda \ln  \left ( \frac{\mu _1}{\mu _2} \right )
+ \lambda \ln  \left ( \frac{\mu _3}{\mu _4} \right )
 + \hat \psi_D,
\end{equation}
with $\mu _3 = -z_2-z + \sqrt{\rho^2 + (z_2+z)^2}$  and   $\mu _4 = -z_1-z +
\sqrt{\rho^2 + (z_1+z)^2}$.
In  the case when the rods just touch
the horizon of the black hole, that is $z_1 = m $,  we have 
\begin{equation}
 \hat \psi =  (1-2\lambda) \hat \psi_s + \lambda \ln  \left ( \frac{\mu _3}{\mu
_2} \right ) 
 + \hat \psi_D.
\end{equation}

The other metric function $\Lambda= \hat \Lambda $ is given by 
\begin{equation}
 \Lambda =   \Lambda_S + \Lambda_{R1} +  \Lambda_{R2}  +  
\Lambda_{D} +  \Lambda_{int}, \label{eq:Lambdasuper}
\end{equation}
where  $\Lambda_{R1}$ and  $\Lambda_{R2}$ are the potential associated with each
rod,   and   by analogy with  (\ref{eq:LSDprola})
\begin{equation} 
 \Lambda_{int}  =  2 \int _{-1}^y \frac{1}{(x^2-y^2)} 
\left [  y ( x^2 - 1 ) \left ( \hat \psi_{R1} +\hat \psi_{R2} +  \hat \psi _D  
\right )_{,x}  + x ( 1 - y^2 )  \left ( 
\hat \psi_{R1} +\hat \psi_{R2} + \hat \psi_D  \right )_{,y} 
\right ] dy.     
\end{equation}
For $z>0$ this integral gives
\begin{equation}
\Lambda_{int}  =   2 \lambda  \ln \left [  \frac{ -m y + z_2 x - \sqrt{r^2 +
(z_2 - z )^2}  }
{m y + z_2 x + \sqrt{r^2 + (z_2 + z )^2  } }     \right] 
-  2 \lambda  \ln \left [  \frac{ x - 1  }{x + 1 }  \right]  
-2 \lambda \ln \left(    \frac{x^2 -1} {x^2 - y^2}     \right ) + \Lambda_{SD},
\end{equation}
being   $\Lambda_{SD}$ the  interaction term   (\ref{eq:LSDcurzon}).

From (\ref{eq:ener})-(\ref{eq:v2super}), the main  physical quantities associated
with the system  are
\begin{eqnarray}
\tilde \epsilon & =&  \frac{ 4 M z_0 e^{ \psi -\Lambda } } {  (\rho^2 +
z_0^2)^{3/2} } \left [F -  (1- 2\lambda) \frac{m }{ \sqrt {\rho^2 + m^2} } 
- \frac{2 \lambda z_2 }{ \sqrt{\rho ^2 + z_2^2} }- 
 \frac{M \rho^2 }{ (\rho^2 + z_0^2)^{3/2}  } \right ] , \\
\tilde p_{\varphi} &=& \frac{ 4 M z_0 e^{ \psi -\Lambda } } { (\rho^2 +
z_0^2)^{3/2} } \left [   (1-2\lambda) \frac{m }{ \sqrt {\rho^2 + m^2} } 
+ \frac{2 \lambda z_2 }{ \sqrt{\rho ^2 + z_2^2} } 
+  \frac{M \rho^2 }{ (\rho^2 + z_0^2)^{3/2} }   \right ] , 
\label{eq:pragn} \\
\text{\sl j}& = &  - \frac{M b}{2 \pi}  e^{2\psi -\Lambda}  \frac{\rho
}{(\rho^2 + z_0^2)^{3/2}} , \label{eq:jagn} \\
v^2 & = &  \frac{F  \left [    (1-2\lambda) \frac{m }{ \sqrt {\rho^2 + m^2} } 
+ \frac{2 \lambda z_2 }{ \sqrt{\rho ^2 + z_2^2} } 
+  \frac{M \rho^2 }{ (\rho^2 + z_0^2)^{3/2} }      \right ] }
 { 1- F \left [    (1-2\lambda) \frac{m }{ \sqrt {\rho^2 + m^2} } 
+ \frac{2 \lambda z_2 }{ \sqrt{\rho ^2 + z_2^2} } 
+  \frac{M \rho^2 }{ (\rho^2 + z_0^2)^{3/2} }   \right ] 
 +  \frac{1}{8 \pi} b^2 f } .  \label{eq:v2agn}
\end{eqnarray}

\begin{figure}
$$
\begin{array}{ccc}
\tilde \epsilon    &  \tilde p_\varphi &  -\text{\sl j} \\
\includegraphics[width=0.3\textwidth]{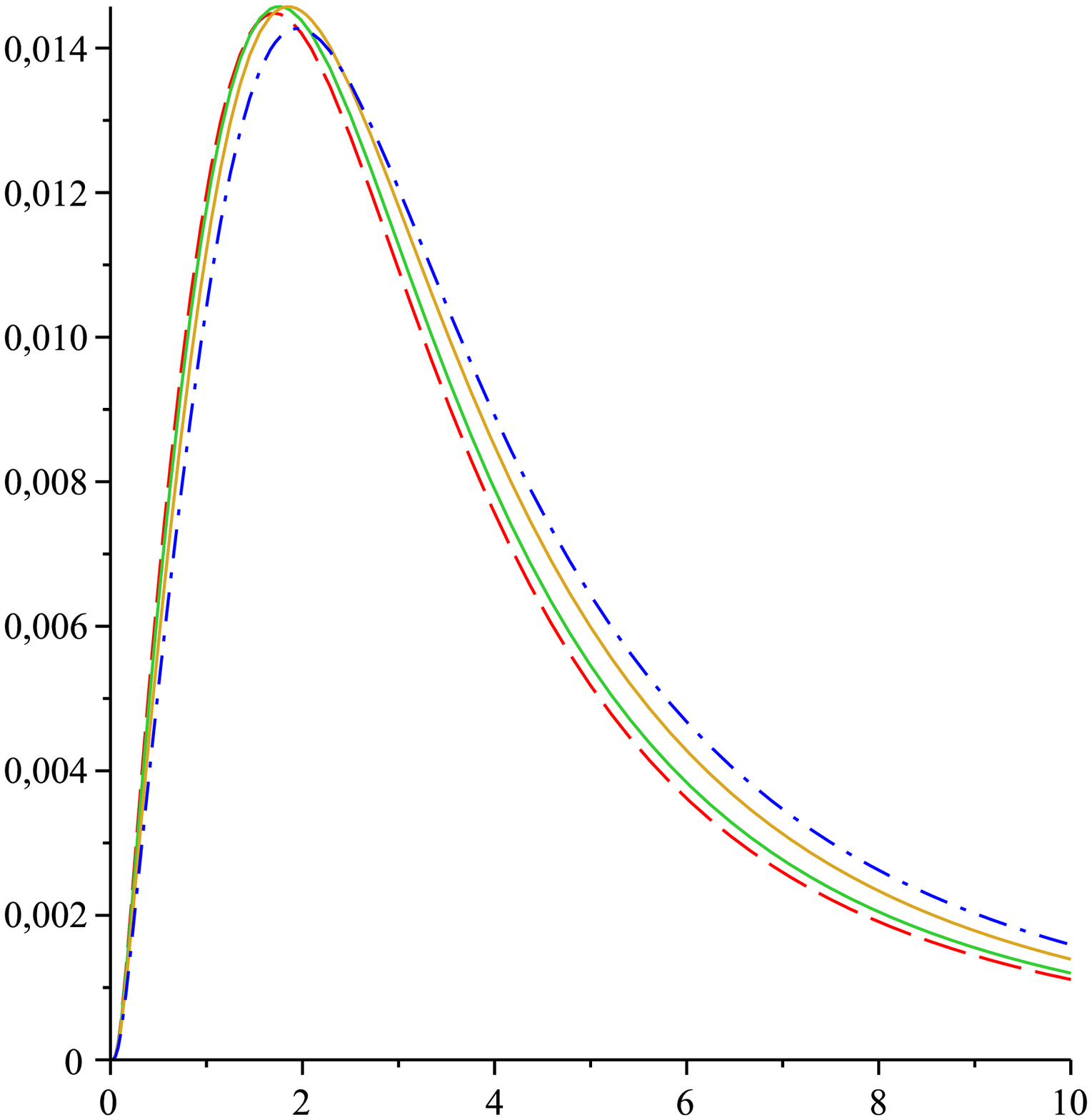} &      
\includegraphics[width=0.3\textwidth]{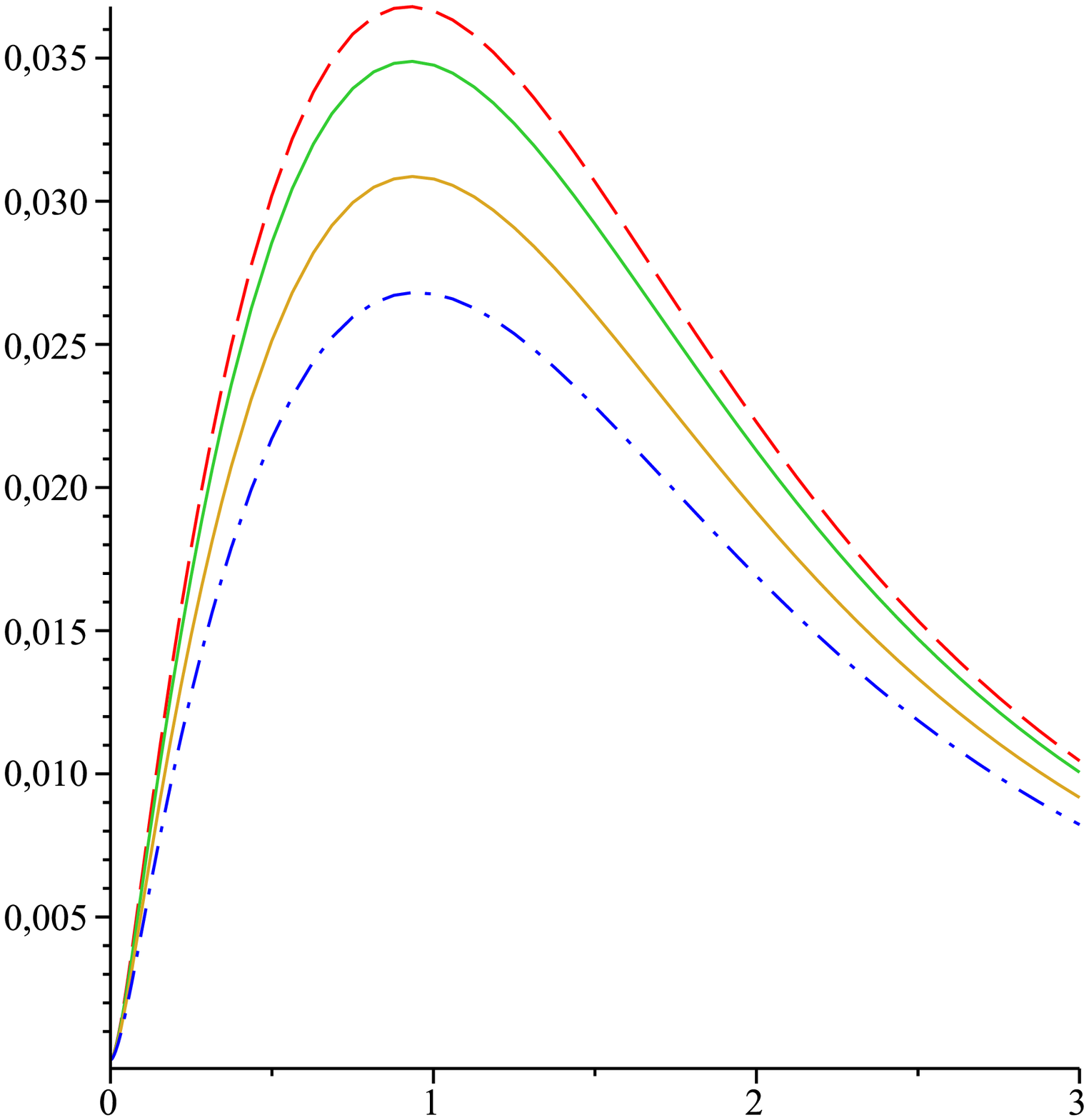} &
\includegraphics[width=0.3\textwidth]{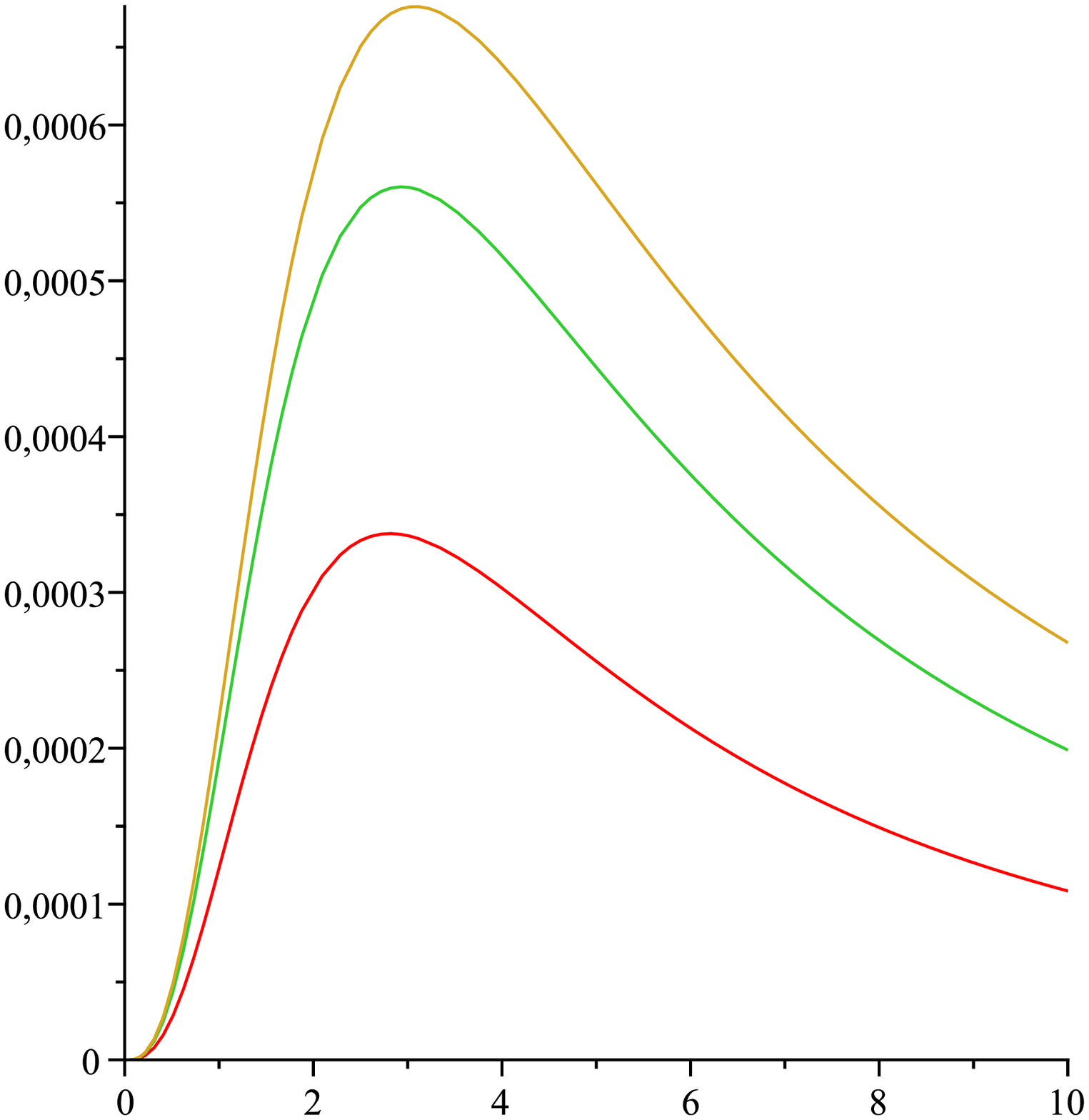}   \\ 
(a)    &  (b) & (c) 
\end{array}
$$	
\caption{$(a)$  The  energy density $\tilde \epsilon$ and   $(b)$ the azimuthal
pressure $\tilde p_\varphi$ for the system composed of a   black hole, 
 a  Kuzmin-Chazy-Curzon  disk and two rods in a magnetic field  
with parameters  $z_0=2$,  $m =0.1$,  $M = 0.2$, $z_2=2$, $\lambda=0.4$,   and
a magnetic field   $b= 0$ (dashed curves), $0.5$,  $1$,
and $1.5$ (dash-dotted  curves), as  functions of $\rho$. $(c)$ The azimuthal electric
current density $\text{\sl j}$ for
$b= 0$ (axis $\rho$), $0.5$,  $1$, and $1.5$ (top curve) and the same value of
other parameters.}
\label{fig:agn1}
\end{figure}

\begin{figure}
$$
\begin{array}{cc}
v^2   &  h^2  \\
\includegraphics[width=0.3\textwidth]{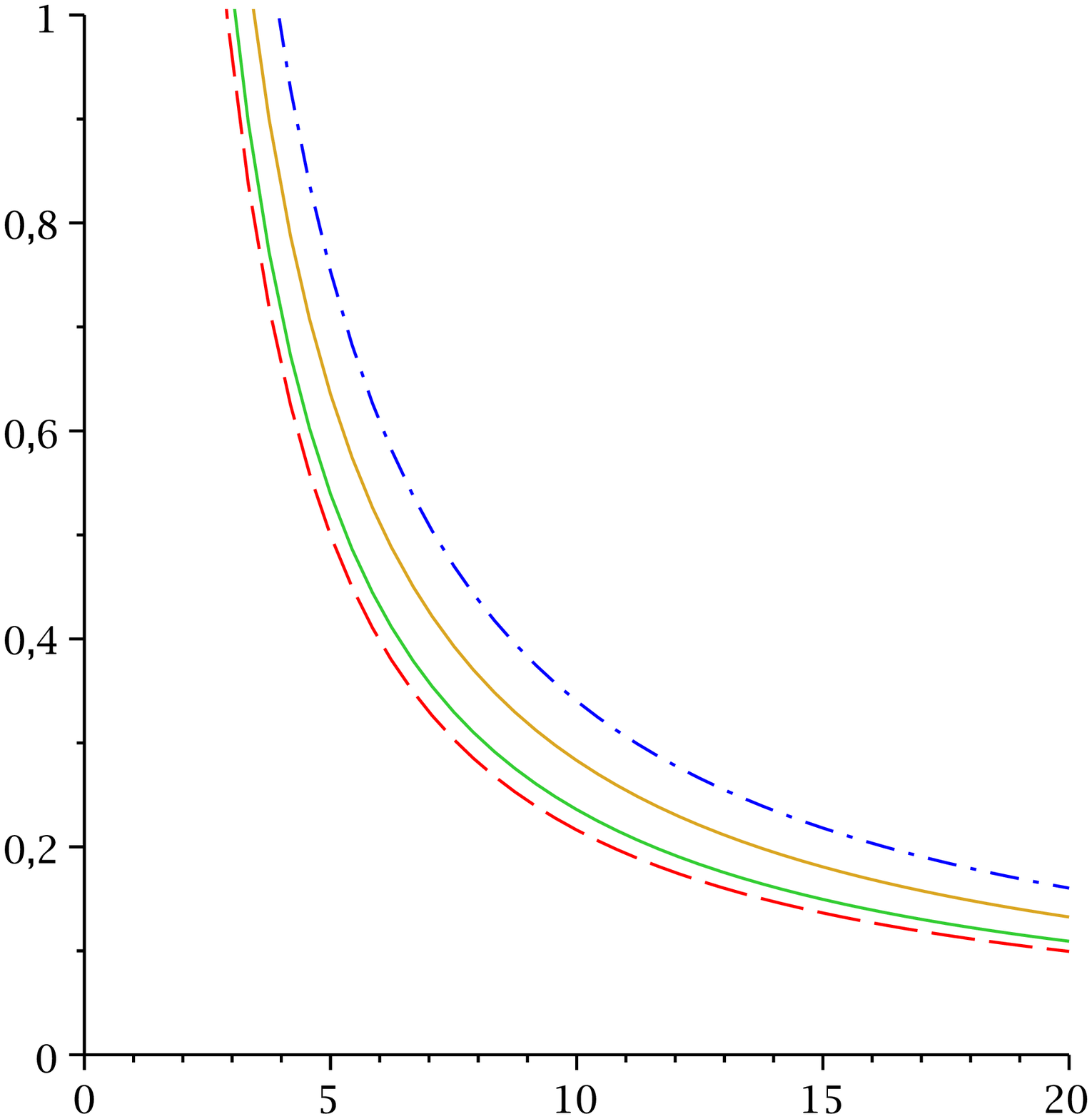} &      
\includegraphics[width=0.3\textwidth]{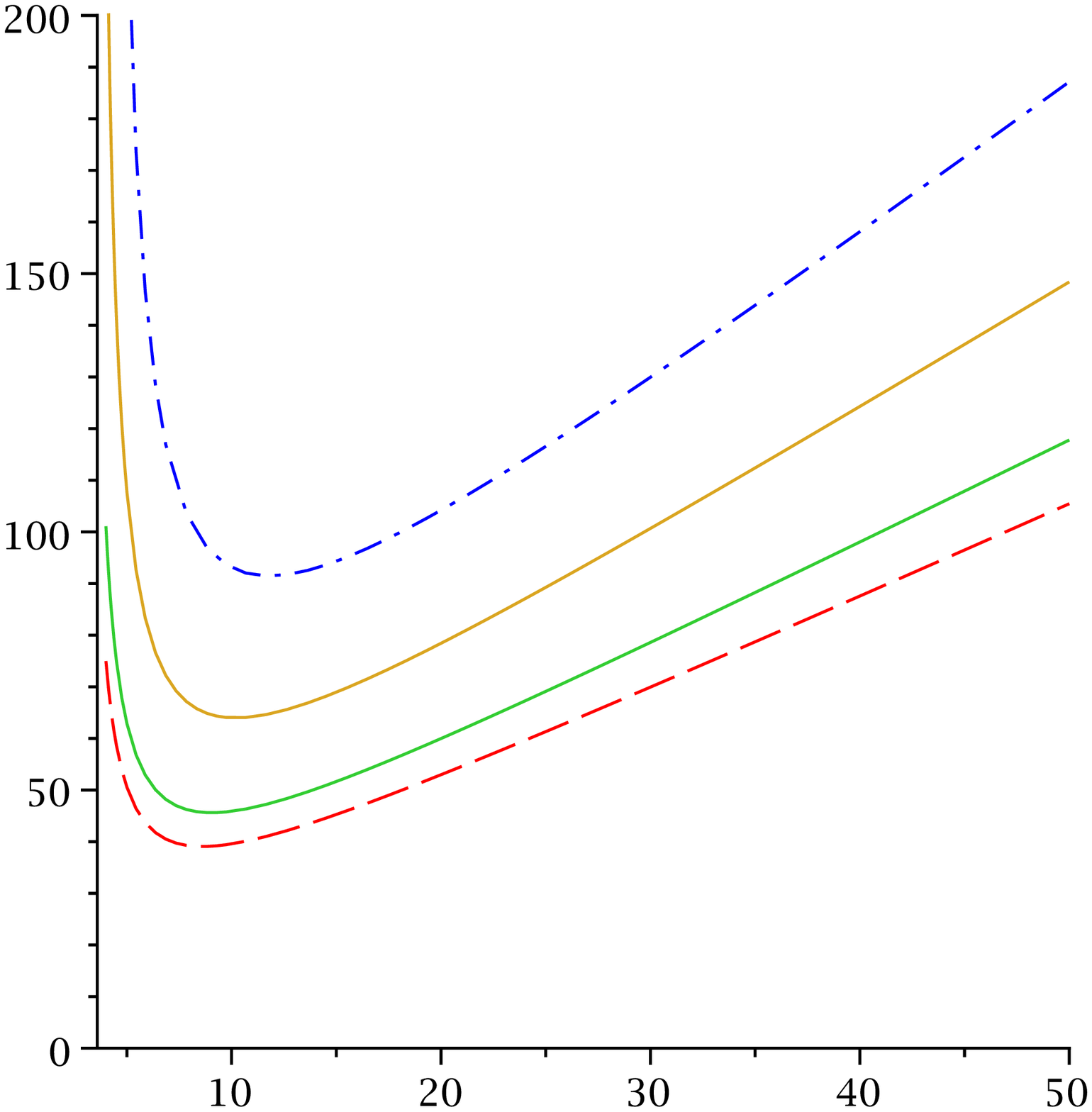}  \\ 
(a)    &  (b)  
\end{array}
$$	
\caption{$(a)$  The speed   $v^2$ of counterrrotation and
$(d)$ the specific angular momentum $h^2$   
for the above AGN model
with    $b= 0$ (dashed curves), $0.5$,  $1$, and $1.5$ (dash-dotted curves) and the same
value of other parameters.}
\label{fig:agn2}
\end{figure}

In Figs.  \ref{fig:agn1} and  \ref{fig:agn2}  we plot   
the  energy density $\tilde \epsilon$, the azimuthal
pressure $ \tilde p_\varphi$,  the azimuthal electric current density $\text{\sl j}$,
the speed   $v^2$ of counterrrotation and
the specific angular momentum $h^2$ as  function of $\rho$   
for   $z_0=2$,  $m =0.1$,  $M = 0.2$, $z_2=2$, $\lambda=0.4$,   and
different  values of magnetic field parameter  $b= 0$, $0.5$,  $1$,
and $1.5$.   We see that  energy density is lowered with increasing magnetic
field near  the center of disk and then is increased, while the pressure 
is lowered in all regions of the disk. 
We also  observer   that  energy density is a positive quantity  in
concordance with  the weak energy condition,  and Eq. (\ref{eq:pragn}) shows 
that 
for $\lambda \leq 1/2$  we   have  pressure 
for all values of parameters.   
The dominant energy condition ($v\leq 1$)  is also  not satisfied  in the
central
region located between the black hole and the photon radius, 
and  increasing  magnetic field   enhances the zone of superluminal  speed. 
We also observer that these structures  present  strong instabilities  inside
the photonic orbit, and that  the magnetic field  increases   the region of
instability. 


\section{Discussion}

The formalism for superposing the field of a  disk and a static black hole in the vacuum
was extended to the case of  magnetized Weyl fields.
Two relativistic models of thin disk  around 
static black hole in presence of a 
magnetic field were presented. 
The first model  is based in the Kuzmin-Chazy-Curzon infinite disk
and the other in the first Morgan-Morgan finite disk. 
We also  considered 
a simple  model of  active galactic nuclei
based on the  superpositions of black hole, a Kuzmin-Chazy-Curzon disk and two
rods, representing the matter of jets, in  presence of magnetic field. 
In the first and three  model  the source of the magnetic  field is
the  surface electric current density  presents on  disk whereas in
second  the source is not planar.
These  solutions were   interpreted as  the exterior gravitational field  of
a black hole and  a finite disk  immersed in a magnetic field. We concluded in this case  that in order to
construct a planar source for the magnetic field is necessary to include the electric field.

In all cases we found values of parameters for which  
energy density is a positive quantity  in
concordance with  the weak energy condition, 
 as well as the stress in
azimuthal  direction (pressure). 
However, the dominant energy condition ($v\leq 1$)  is  not satisfied in all regions of the disks 
and the  presence 
magnetic field  enhances the zones of superluminal  speed. 
Such   behavior is also  observed in the second models   when the electric field is included. 
We also  see   that  the magnetic field  increases   the regions of
instability. 

Finally, in order to construct a such  system that satisfies all the energy
conditions, 
a model of a black hole  surrounded by a  disk with an inner edge
in presence of magnetic field   is being investigated.

\begin{acknowledgments}
One of us (A.C.G-P.) wants to thank  COLCIENCIAS, TWAS and Conacyt
for financial support. 
\end{acknowledgments}

\appendix*
\section{}

The components of the energy-momentum tensor of the disk $Q_{a}^{b}$  can be 
obtained by  integration of the field equations (\ref{eq:EMdisk}) writing 
$T_{ab}$ as  (\ref{eq:Tabdisk})  \cite{CHGS}   
\begin{equation}
\int _{z=0_-}^{z=0_+} R_{ab} dz = 8 \pi ( \int _{z=0_-}^{z=0_+}  T^{\text{elm}}_{ab}dz
+   \int _{z=0_-}^{z=0_+} Q_{ab} \ \delta (z) dz).   \label{eq:intRab}
\end{equation}
For metric (\ref{eq:met}), the  nonzero components of $T^{\text{elm}}_{ab}$ are   
\begin{subequations}\begin{eqnarray}
T^{\text{elm}}_{00} & = & \frac { e^{ 2 (3 \psi - \Lambda) } }   {8 \pi \rho^2}
(   A_{,\rho}^2 +   A_{,z}^2  ), \\
T^{\text{elm}}_{11} & = & \frac  {1} {8 \pi}  e^{ 2 (\psi - \Lambda) }  (  
A_{,\rho}^2 +   A_{,z}^2  ),  \\
T^{\text{elm}}_{22} & = & - T^{\text{elm}}_{33} = \frac { e^{ 2 \psi  } }  {8
\pi  \rho^2} (   A_{,\rho}^2 -    A_{,z}^2  ), \\
T^{\text{elm}}_{23} & = & \frac { e^{ 2 \psi  } }  {4 \pi \rho^2}    A_{,\rho}
A_{,z},  
\end{eqnarray}\end{subequations}
and for magnetized Weyl  fields (Eq. (\ref{eq:A})) give us  
\begin{subequations}\begin{eqnarray}
T^{\text{elm}}_{00} & = & \frac {b^2 }{8 \pi } e^{ 2 (3 \psi - \Lambda) }  
(  \hat \psi_{,\rho}^2 + \hat \psi_{,z}^2  ), \\
T^{\text{elm}}_{11} & = & \frac  {b^2 } {8 \pi} \rho^2  e^{ 2 (\psi - \Lambda) }
(  \hat \psi_{,\rho}^2 + \hat \psi_{,z}^2  ),  \\
T^{\text{elm}}_{22} & = & - T^{\text{elm}}_{33} = 
\frac {b^2 }{8 \pi } e^{ 2 \psi  }  ( \hat \psi_{,z}^2 - \hat \psi_{,\rho}^2 )  \\
T^{\text{elm}}_{23} & = & -  \frac {b^2 }{4 \pi }  e^{ 2 \psi  }   
\hat \psi_{,\rho}   \hat \psi_{,z} .  
\end{eqnarray}\end{subequations}

Now the continuity of the metric functions across the disk implies that they are even functions of $z$
and the discontinuity of its first derivatives in the direction  normal to the disk  means
that these  are odd functions of $z$. Thus  $\hat \psi_{,z}$ is odd and its square  is even and hence
continuous across the disk. It also follows that $\hat \psi_{, \rho}$ is continuous.  Accordingly the terms  
\begin{equation}
\int _{z=0_-}^{z=0_+}  T^{\text{elm}}_{00} dz  
= \int _{z=0_-}^{z=0_+}  T^{\text{elm}}_{11} dz = \int _{z=0_-}^{z=0_+} 
T^{\text{elm}}_{22} dz
= \int _{z=0_-}^{z=0_+}  T^{\text{elm}}_{33} dz=0 
\end{equation}
are zero   due to the continuity of the metric, $ \hat \psi_{,\rho}^2$ and $ \hat \psi_{,z}^2$
across the disk. For the other component we have
\begin{equation}
\int _{z=0_-}^{z=0_+}   T^{\text{elm}}_{23}    dz = 
-  \frac {b^2 }{4 \pi }  e^{ 2 \psi  }   \hat \psi_{,\rho} 
 \int _{z=0_-}^{z=0_+}  \hat \psi_{,z}    dz =   
 -  \frac {b^2 }{4 \pi }  e^{ 2 \psi  }   \hat \psi_{,\rho}[ \hat \psi] =0,
\end{equation}
which also vanishes because continuity of the metric and  $\hat \psi_{, \rho}$ across the
disk.
Therefore,  the first term  on right hand side  of (\ref{eq:intRab})
is zero and in consequence $Q_a^b$ has the same form that the vacuum
(\ref{eq:Qab}).  

On the other hand,  on the disk the Maxwell equations  $\partial_ b \bar F^{ab}
= 4 \pi \bar
J^a$, where `bar' denotes multiplication by $\sqrt{-g}$,   are given by 
\begin{equation}
-4 \pi \bar j_\varphi \delta(z) =   \partial _z  ( \bar  g^{zz}  A_{,z} ) 
+ \partial _\rho ( \bar  g^{\rho \rho}   A_{,\rho}  ),
\end{equation}
but again  using (\ref{eq:A}) we have
\begin{equation}
-4 \pi \bar j_\varphi \delta(z) = 
b \rho \partial _z  ( \bar  g^{zz} \hat \psi _{,\rho} ) 
- b  \partial _\rho (  \rho  \bar  g^{\rho \rho}    \hat \psi_{,z}  ). 
\end{equation}
Integrating through the disk one finds that the electric current density   is  zero on the disk. In fact 
\begin{eqnarray}
-4 \pi \bar j_\varphi  & = &  b \rho  \int _{z=0_-}^{z=0_+}  \partial _z ( \bar g^{zz} \hat \psi _{,\rho}  ) dz
- b  \int _{z=0_-}^{z=0_+} \partial _{\rho} (  \rho  \bar  g^{\rho \rho}    \hat \psi_{,z}  ) dz 
\nonumber  \\
&=&  b  \rho  ( \bar  g^{zz}  \hat \psi _{,\rho}) |_{z=0_-}^{z=0_+} 
-  b  \partial _{\rho}  (  \rho \bar  g^{\rho \rho}   \int _{z=0_-}^{z=0_+}  \hat \psi_{,z} dz ) 
\nonumber \\
& = & b  \rho  [ \bar g^{zz}  \hat \psi _{,\rho} ] 
- b \partial _{\rho} (  \rho \bar  g^{\rho \rho} [  \hat \psi ]  ) 
\nonumber \\   
&=& 0, 
\end{eqnarray}
where the   terms on the right hand side vanish  due to the  continuity of the
metric and  $\hat \psi _{,\rho}$  across the disk. 
Thus, if the  gravitational field is continuous across plane $z=0$ but
its normal derivative
is discontinuous, the magnetostatic solutions    discussed in IV (B)   
can be  interpreted  as  the superposition of a black hole  
and  a finite disk  immersed in a magnetic field.


\newpage




\begin{thebibliography}{9999}



\bibitem{Kundt} W. Kundt (Ed.), Jets from Stars and Galactic Nuclei,
Proceedings,
Lecture Notes in Physics, 471, Springer, 1996.

\bibitem{Krolik} J. H. Krolik, Active Galactic Nuclei: from the Central Black
Hole to
the Galactic Environment, Princeton University Press, Princeton, New
Jersey, 1999.

\bibitem{Blandford} R. D. Blandford, Prog. Theor. Phys. Supp. {\bf 143}, 182
(2001).


\bibitem{BZ} R. D.  Blandford and R. L.  Znajek, MNRAS  {\bf 179},  433 (1977).


\bibitem{Zakharov} A. F. Zakharov, N. S. Kardashev, V. N. Lukash, S. V. Repin, 
MNRAS  {\bf 342},  1325 (2003)

\bibitem{Silantev} N. A. Silant'ev, Yu. N. Gnedin, S. D. Buliga, M. Yu.
Piotrovich, T. M. Natsvlishvili,
arXiv:1203.2763v1.  


\bibitem{Liu} Y.T. Liu, S. L. Shapiro and B. C. Stephens, Phys. Rev. D
76, 084017 (2007).

\bibitem{Akiyama} S. Akiyama et al, ApJ 584, 954 (2003).

\bibitem{Vlemming} W. H. T. Vlemmings, P. J. Diamond and H. Imai, Nature
440, 58 (2006).

\bibitem{Meier} D. L. Meier, S. Koide and Y. Uchida, Science 291, 84 (2001).

\bibitem{Ustyugova} G. V. Ustyugova et al, ApJ 516, 221 (1999).

\bibitem{Krasnopolsky} R. Krasnopolsky, Z. Y. Li and R. D. Blandford, ApJ 595,
631 (2003).


\bibitem{BS} W. A. Bonnor and A. Sackfield, Commun. Math.   Phys. {\bf 8}, 338
(1968).

\bibitem{MM1} T. Morgan and L. Morgan, Phys. Rev.  {\bf 183},  1097 (1969).

\bibitem{MM2} L. Morgan and T. Morgan, Phys. Rev. D  {\bf 2},  2756 (1970).



\bibitem{LP} D. Lynden-Bell and S. Pineault, Mon. Not. R.  Astron. Soc. {\bf
185}, 679 (1978). \label{bib:LP} 

\bibitem{CHGS} A. Chamorro, R. Gregory, and J. M. Stewart, Proc. R. Soc. London
{\bf A413}, 251 (1987).

\bibitem{LO} P.S. Letelier and S. R. Oliveira, J. Math.  Phys.  {\bf 28}, 165
(1987).

\bibitem{LEM} J. P. S. Lemos, Class. Quantum Grav. {\bf 6}, 1219 (1989).

\bibitem{BLK} J. Bi\u{c}\'{a}k, D. Lynden-Bell, and J.  Katz,  Phys. Rev. D {\bf
47}, 4334 (1993).

\bibitem{BLP} J. Bi\u{c}\'{a}k, D. Lynden-Bell, and C.  Pichon, Mon. Not. R.
Astron. Soc. {\bf 265}, 126 (1993).

\bibitem{GE} G. A. Gonz\'alez and O. A. Espitia,  Phys. Rev. D  {\bf 68}, 104028
(2003). 



\bibitem{BL} J. Bi\u{c}\'ak and T. Ledvinka, Phys. Rev.  Lett. {\bf 71}, 1669 
(1993).

\bibitem{GL2} G. A. Gonz\'alez and P. S. Letelier, Phys.  Rev.  D {\bf 62},
064025 (2000).


\bibitem{LL1} J. P. S. Lemos and P. S. Letelier, Class.  Quantum Grav. {\bf
10}, L75 (1993).  

\bibitem{LL2} J. P. S. Lemos and P. S. Letelier, Phys. Rev. D  {\bf 49},  5135
(1994).  

\bibitem{LV-AGN} D. Vogt and P. S. Letelier,  Phys. Rev. D {\bf 71}, 044009
(2005) 


\bibitem{LBZ} T. Ledvinka, J. Bi\u{c}\'{a}k, and M.  \u{Z}ofka, in {\it
Proceeding of 8th Marcel-Grossmann  Meeting in General Relativity}, edited by
T. Piran  (World  Scientific, Singapore, 1999) 

\bibitem{KBL} J. Katz, J. Bi\u{c}\'ak, and D. Lynden-Bell, Class. Quantum Grav.
{\bf 16}, 4023 (1999).

\bibitem{LET1} P. S. Letelier, Phys. Rev. D {\bf 60},  104042  (1999). 

\bibitem{GG1} G. Garc\'\i a R. and G. A. Gonz\'alez, Phys.  Rev. D  {\bf 69},
124002 (2004).

\bibitem{GG-CRIS} C. H. Garc\'{\i}a-Duque and G.  Garc\'{\i}a-Reyes, Gen.
Relativ. Gravit. {\bf 43}, 11, 3001 (2011).

\bibitem{Guti1}  A. C. Guti\'errez-Pi\~neres and G. A. Gonz\'alez, Int. J. Theor. Phys. {\bf 43}, 6, 1737 (2012).

\bibitem{Guti2} N. G\"urlebeck, J.  Bi\u{c}\'{a}k  and A. C. Guti\'errez-Pi\~neres, Phys.  Rev. D  {\bf 83},
124023 (2011).

\bibitem{Guti3}  N. G\"urlebeck, J.  Bi\u{c}\'{a}k  and A. C. Guti\'errez-Pi\~neres, Gen. Relativ.
Gravit. {\bf 43},  3301 (2011).   	

\bibitem{AN-GUI-QUE}  A. C. Guti\'errez-Pi\~neres, G. A.  Gonz\'alez, and H. Quevedo, Phys. Rev.  D 
{\bf 87}, 044010 (2013).

\bibitem{AN-QUE} A. C. Guti\'errez-Pi\~neres, C. S. Lopez-Monsalvo, H. Quevedo,   arXiv:1306.6591 


\bibitem{GON-OMAR}G.  Garc\'{\i}a-Reyes and Omar A. Espitia,  arXiv:1307.2281.




\bibitem{E1} F. J. Ernst. Phys. Rev. {\bf 167}, 1175 (1968).

\bibitem{E2} F. J. Ernst. Phys. Rev. {\bf 168}, 1415 (1968).


\bibitem{KRAMER} D. Kramer, H. Stephani, E. Herlt, and  M. McCallum, 
{\it Exact Solutions of Einsteins's  Field Equations} (Cambridge University
Press, Cambridge, England, 1980).



\bibitem{PH} A. Papapetrou and A. Hamouni, Ann. Inst. Henri Poincar\'e {\bf 9},
179 (1968)

\bibitem{LICH} A. Lichnerowicz, C.R. Acad. Sci. {\bf 273}, 528 (1971) 

\bibitem{TAUB} A. H. Taub, J. Math. Phys. {\bf 21}, 1423 (1980)

\bibitem{IS1} E. Israel, Nuovo Cimento {\bf 44B}, 1 (1966)

\bibitem{IS2} E. Israel, Nuovo Cimento {\bf 48B}, 463 (1967)



\bibitem{RGK} V. C. Rubin, J. A. Graham and J. D. P Kenney. Ap. J. {\bf 394},
L9, (1992).

\bibitem{RFF} H. Rix, M. Franx, D. Fisher and G. Illingworth. Ap. J. {\bf 400},
L5, (1992).

\bibitem{BER} F. Bertola {\it et al}. Ap. J. {\bf 458}, L67 (1996).

\bibitem{STRUCK} C. Struck, Phys. Rep. {\bf 321}, 1 (1999).

\bibitem{CBG} R. Ciri, D. Bettoni, and G. Galletta, 
Nature {\bf 375}, 661 (1995).


\bibitem{KLE3} C. Klein, Phys. Rev.  D {\bf 65}, 084029  (2002).

\bibitem{GGR-Inter} G. Garc\'\i a-Reyes  and G. A. Gonz\'alez,  Int. J. Mod. Phys. D  {\bf 18}, 9,
1461  (2009).  


\bibitem{RAYL} Lord Rayleigh, 1917, Proc. R. Soc. London A, 93, 148

\bibitem{FLU} L. D. Landau and E.M. Lifshitz, {\it Fluid 
Mechanics}(Addison-Wesley, Reading, MA, 1989).

\bibitem{LETSTAB} P. S.  Letelier,  Phys. Rev. D {\bf 68}, 104002 (2003).






\bibitem{KUZMIN} G. G. Kuzmin 1956,  Astron. Zh.,  33, 27 (1956) 

\bibitem{TOOMRE} A. Toomre, Ap. J., 138, 385  (1962)  


\bibitem{CH} Chazy J 1924 Bull. Soc. Math., France 52 17.

\bibitem{C} Curzon H E J 1924 Proc. London Math. Soc. 23 477. 


\bibitem{FOSSIL} J. Braithwaite and  H. C. Spruit, Nature 431, 819 
 (2004). 

\end{thebibliography}
\end{document}